\documentclass[longauth]{aa} % for the long lists of affiliations 

\usepackage{graphicx}
\usepackage{txfonts}
\usepackage{longtable}
\usepackage[normalem]{ulem}

\usepackage{xcolor}

\usepackage{subfig}
\usepackage{threeparttable}
\usepackage[colorlinks=true, citecolor=blue]{hyperref}

\usepackage{natbib}
\bibpunct{(}{)}{;}{a}{}{,} % to follow the A&A style

\begin{document} 

\title{Overdense fireworks in GOODS-N: Unveiling a record number of massive dusty star-forming galaxies at z$\sim$5.2 with the N2CLS}

   \author{G. Lagache \inst{\ref{LAM}}\thanks{\email{guilaine.lagache@lam.fr}}
     \and M.~Xiao \inst{\ref{UniGe}}
     \and A.~Beelen \inst{\ref{LAM}} 
     \and S.~Berta \inst{\ref{IRAMF}}
     \and L.~Ciesla \inst{\ref{LAM}} 
     \and R.~Neri \inst{\ref{IRAMF}} 
     \and R.~Pello \inst{\ref{LAM}} 
     \and R.~Adam \inst{\ref{OCA}}
     \and  P.~Ade \inst{\ref{Cardiff}}
     \and  H.~Ajeddig \inst{\ref{CEA}}
     \and  S.~Amarantidis \inst{\ref{IRAME}}
     \and  P.~Andr\'e \inst{\ref{CEA}}
     \and  H.~Aussel \inst{\ref{CEA}}
     \and  A.~Beno\^it \inst{\ref{Neel}}
     \and M.~Béthermin \inst{\ref{UniStra}}
     \and L.-J.~Bing \inst{\ref{Sussex}}
     \and  A.~Bongiovanni \inst{\ref{IRAME}}
     \and  J.~Bounmy \inst{\ref{LPSC}}
     \and  O.~Bourrion \inst{\ref{LPSC}}
     \and  M.~Calvo \inst{\ref{Neel}}
     \and  A.~Catalano \inst{\ref{LPSC}}
     \and  D.~Ch\'erouvrier \inst{\ref{LPSC}}
     \and  U.~Chowdhury \inst{\ref{Neel}}       
     \and  M.~De~Petris \inst{\ref{Roma}}
     \and  F.-X.~D\'esert \inst{\ref{IPAG}}
     \and  S.~Doyle \inst{\ref{Cardiff}}
     \and  E.~F.~C.~Driessen \inst{\ref{IRAMF}}
     \and  G.~Ejlali \inst{\ref{Teheran}}
     \and  A.~Ferragamo \inst{\ref{Roma}}
     \and  A.~Gomez \inst{\ref{CAB}} 
     \and  J.~Goupy \inst{\ref{Neel}}
     \and  C.~Hanser \inst{\ref{CPPM}}
     \and  S.~Katsioli \inst{\ref{AthenObs}, \ref{AthenUniv}}
     \and  F.~K\'eruzor\'e \inst{\ref{Argonne}}
     \and  C.~Kramer \inst{\ref{IRAMF}}
     \and  B.~Ladjelate \inst{\ref{IRAME}} 
     \and  S.~Leclercq \inst{\ref{IRAMF}}
     \and  J.-F.~Lestrade \inst{\ref{LUX}}
     \and  J.~F.~Mac\'ias-P\'erez \inst{\ref{LPSC}}
     \and  S.~C.~Madden \inst{\ref{CEA}}
     \and  A.~Maury \inst{\ref{Barcelona1}, \ref{Barcelona2}, \ref{CEA}}
     \and  F.~Mayet \inst{\ref{LPSC}}
     \and  A.~Monfardini \inst{\ref{Neel}}
     \and  A.~Moyer-Anin \inst{\ref{LPSC}}
     \and  M.~Mu\~noz-Echeverr\'ia \inst{\ref{IRAP}}
     \and  I.~Myserlis \inst{\ref{IRAME}}
     \and P.~Oesch \inst{\ref{UniGe},\ref{dawn},\ref{Copenhagen}}
     \and  A.~Paliwal \inst{\ref{Roma2}}
     \and  L.~Perotto \inst{\ref{LPSC}}
     \and  G.~Pisano \inst{\ref{Roma}}
     \and  N.~Ponthieu \inst{\ref{IPAG}}
     \and  V.~Rev\'eret \inst{\ref{CEA}}
     \and  A.~J.~Rigby \inst{\ref{Leeds}}
     \and  A.~Ritacco \inst{\ref{LPSC}}
     \and  H.~Roussel \inst{\ref{IAP}}
     \and  F.~Ruppin \inst{\ref{IP2I}}
     \and  M.~S\'anchez-Portal \inst{\ref{IRAME}}
     \and  S.~Savorgnano \inst{\ref{LPSC}}
     \and  K.~Schuster \inst{\ref{IRAMF}}
     \and  A.~Sievers \inst{\ref{IRAME}}
     \and  C.~Tucker \inst{\ref{Cardiff}}
     \and  R.~Zylka \inst{\ref{IRAMF}}
     }
   \institute{Aix Marseille Univ, CNRS, CNES, LAM (Laboratoire d'Astrophysique de Marseille), Marseille, France
     \label{LAM}
     \and
     Department of Astronomy, University of Geneva, Chemin Pegasi 51, 1290 Versoix, Switzerland
     \label{UniGe}
     \and
    Institut de Radioastronomie Millim\'etrique (IRAM), 300 rue de la Piscine, 38400 Saint-Martin-d'H{\`e}res, France
     \label{IRAMF}
    \and
    Universit\'e C\^ote d'Azur, Observatoire de la C\^ote d'Azur, CNRS, Laboratoire Lagrange, France 
     \label{OCA}
    \and
     School of Physics and Astronomy, Cardiff University, Queen’s Buildings, The Parade, Cardiff, CF24 3AA, UK 
     \label{Cardiff}
     \and
     Université Paris Cité, Université Paris-Saclay, CEA, CNRS, AIM, F-91191 Gif-sur-Yvette, France
     \label{CEA}
     \and
     Institut de Radioastronomie Millim\'etrique (IRAM), Avenida Divina Pastora 7, Local 20, E-18012, Granada, Spain
     \label{IRAME}     
     \and
     Institut N\'eel, CNRS, Universit\'e Grenoble Alpes, France
     \label{Neel}
     \and
    Universit\'e de Strasbourg, CNRS, Observatoire astronomique de Strasbourg, UMR 7550, 67000 Strasbourg, France
     \label{UniStra}
    \and
    Astronomy Centre, Department of Physics and Astronomy, University of Sussex, Brighton BN1 9QH
     \label{Sussex}
     \and
      Univ. Grenoble Alpes, CNRS, Grenoble INP, LPSC-IN2P3, 53, avenue des Martyrs, 38000 Grenoble, France
     \label{LPSC}
     \and 
     Dipartimento di Fisica, Sapienza Universit\`a di Roma, Piazzale Aldo Moro 5, I-00185 Roma, Italy
     \label{Roma}
     \and
     Univ. Grenoble Alpes, CNRS, IPAG, 38000 Grenoble, France 
     \label{IPAG}
     \and
     Institute for Research in Fundamental Sciences (IPM), School of Astronomy, Tehran, Iran
     \label{Teheran}
     \and
     Centro de Astrobiolog\'ia (CSIC-INTA), Torrej\'on de Ardoz, 28850 Madrid, Spain
     \label{CAB}
     \and
     Aix Marseille Univ, CNRS/IN2P3, CPPM, Marseille, France
     \label{CPPM}
     \and
     National Observatory of Athens, Institute for Astronomy, Astrophysics, Space Applications and Remote Sensing, Ioannou Metaxa
     and Vasileos Pavlou GR-15236, Athens, Greece
     \label{AthenObs}
     \and
     Department of Astrophysics, Astronomy \& Mechanics, Faculty of Physics, University of Athens, Panepistimiopolis, GR-15784
     Zografos, Athens, Greece
     \label{AthenUniv}
     \and
     High Energy Physics Division, Argonne National Laboratory, 9700 South Cass Avenue, Lemont, IL 60439, USA
     \label{Argonne}
     \and  
     LUX, Observatoire de Paris, PSL Research University, CNRS, Sorbonne Universit\'e, UPMC, 75014 Paris, France  
     \label{LUX}
     \and
     Institute of Space Sciences (ICE), CSIC, Campus UAB, Carrer de Can Magrans s/n, E-08193, Barcelona, Spain
     \label{Barcelona1}
     \and
     ICREA, Pg. Lluís Companys 23, Barcelona, Spain
     \label{Barcelona2}
     \and
     IRAP, CNRS, Université de Toulouse, CNES, UT3-UPS, (Toulouse), France
     \label{IRAP}
    \and
    Cosmic Dawn Center (DAWN)
    \label{dawn}
    \and
    Niels Bohr Institute, University of Copenhagen, Jagtvej 128, DK-2200 Copenhagen N, Denmark
    \label{Copenhagen}
    \and
     Dipartimento di Fisica, Universit\`a di Roma ‘Tor Vergata’, Via della Ricerca Scientifica 1, I-00133 Roma, Italy        
     \label{Roma2}
     \and
     School of Physics and Astronomy, University of Leeds, Leeds LS2 9JT, UK
     \label{Leeds}
     \and    
     Institut d'Astrophysique de Paris, CNRS (UMR7095), 98 bis boulevard Arago, 75014 Paris, France
     \label{IAP}
     \and
     University of Lyon, UCB Lyon 1, CNRS/IN2P3, IP2I, 69622 Villeurbanne, France
     \label{IP2I}
     \and
     University Federico II, Naples, Italy
     \label{Naples}
  }
          
\titlerunning{N2CLS overdense fireworks in GOODS-N}

   \date{Received 13 June 2025; accepted 12 November 2025}

  \abstract{ High-density environments, such as early galaxy overdensities, play a critical role in hierarchical structure formation and galaxy evolution, providing an ideal setting for accelerated galaxy growth. The GOODS-N overdensity at z$\simeq$5.2 has long been known, but its full extent and richness have only recently been revealed by JWST observations. It is highly elongated along the line of sight, spanning nearly 30\,cMpc. We investigated its dusty galaxy population using the NIKA2 Cosmological Legacy Survey (N2CLS). Within this overdensity, we identify five luminous dusty starbursts that are spectroscopically confirmed, along with three additional candidates supported by robust photometric redshifts. Three of the spectroscopically confirmed galaxies (N2GN\_1\_01, 06, and 23, known as GN10, HDF850.1, and S3, respectively) had already been recognised as members of this exceptional structure. We report the discovery of N2GN\_1\_13 at z$_{\rm spec}$=5.182, a massive dusty star-forming galaxy that we confirmed as part of the overdensity through targeted NOEMA follow-up observations of the N2CLS. In addition, by combining our analysis with JWST/FRESCO data, we identified another dusty galaxy at  z$_{\rm spec}$=5.201 (N2GN\_1\_61).
  The eight dusty galaxies are massive (with a median stellar mass of $\sim9\times$10$^{10}$\,M$_{\odot}$) and highly obscured (with a median A$_V$ of $\sim$3.3) and caught in a short-lived yet extreme starburst phase at z$\sim$5.2. Their high star formation rates (with a median of $\sim$590\,M$_{\odot}$\,yr$^{-1}$), efficient baryon to stellar mass conversion ($\epsilon_{\star}>$20\% for 75\% of the sample), and substantial gas reservoirs and dust content suggest rapid evolution and imminent quenching. Six of these galaxies reside in overdense filaments; the remaining two may trace new distinct structures, which will have to be spectroscopically confirmed. These few dusty galaxies dominate the star formation within the overdensity, contributing more than the numerous H$_{\alpha}$ emitters, and surpassing the cosmic average star formation rate density for this epoch. The presence of numerous very massive, dusty, and intensely star-forming galaxies at z$\sim$5.2 shows that rapid stellar and dust mass assembly was already underway within the first billion years of cosmic history in overdense environments. Their properties, likely driven by efficient gas inflows along cosmic filaments in protocluster regions, suggest an accelerated evolution that current models and simulations have difficulty reproducing.}


   \maketitle

\section{Introduction}
At high redshifts (z$\gtrsim$4), and despite recent progress, the contribution of dust-obscured galaxies to the young Universe remains quite uncertain \citep[e.g.][]{Traina2024}. Recent Atacama Large Millimetre/sub-millimetre Array (ALMA) surveys (such as ALPINE and REBELS) show that dust-obscured (far-IR and millimetre) and unobscured (UV rest-frame) star formation could have made a similar contribution to the cosmic budget even at z$\sim$6 \citep[e.g.][]{Gruppioni2020, Fudamoto2020, Khusanova2021, Barrufet2023, fujimoto2024}. 
In addition, recent JWST observations have reported candidate heavily obscured sources at z$\gtrsim$7 \citep{Akins2023, Rodighiero2023, gandolfi25,martis2025}, although spectroscopic confirmation and direct detection of the thermal dust continuum are still lacking.
These studies are not based on volume-complete surveys and/or rely on complex corrections and selections. The population of luminous and dusty galaxies at z$>$4 is largely unconstrained, and accurately quantifying the total star formation rate density (SFRD) at z$\gtrsim$4 from a well-controlled volume-limited survey remains an ambitious goal. One of the main contributors to the obscured SFRD at such high redshifts are millimetre-bright, ultra-massive, and optically dark galaxies\footnote{Such galaxies are not detected in the deepest HST images (with typical 5-sigma depth of 27\,mag for H band) but are detected with IRAC, and now routinely by the JWST.}. A recent JWST study based on three millimetre-bright sources \citep{Xiao2024} at z$\sim$5-6 shows that they account for $\sim$20\,\% of the SFRD derived from rest-frame far-UV to optical observations and corrected for dust attenuation at $z\sim5.8$. These three galaxies are also ultra-massive (log(M$_{\star}$/M$_{\odot})>$11.0) and would require, on average, about 50\% of the baryons in their halos to be converted into stars. This is two to three times higher than even the most efficient galaxies at later times \citep[][and references therein]{Xiao2024}. 

The two ultra-massive galaxies in the great observatories origins deep surveys north (GOODS-N) field (S3 and GN10) are located in a large-scale structure in the process of formation \citep{Herard-Demanche2024}, which also contains the well known HDF850.1 galaxy. HDF850.1 is forming in one of the richest environments identified to date at z$>$5, with $\sim$100 z=5.17-5.20 galaxies (\citealt{Herard-Demanche2024, Sun2024}; see also e.g. \citealt{calvi2021}). 
Even before the extent of this structure was known, the redshifts of HDF850.1 and GN10 alone already suggested a significant overdensity of dusty star-forming galaxies at z$>$5 in this field compared to expectations, though it remained unclear whether the individual substructures traced by these sources were physically connected \citep[e.g.][]{Riechers2020}.

Observations suggest that such (massive) overdensities are not rare at $z\sim5$ \citep[e.g.][]{guaita2022, helton2024}. They are likely progenitors of today's massive galaxy clusters. Understanding their distribution and properties shortly after the epoch of reionisation helps bridge observations of the early Universe with modern-day structures. Given the existence of a similar galaxy overdensity in both the GOODS-N and GOODS-S fields, \cite{Sun2024} suggest that 50\%$\pm$20\% of the cosmic star formation at z = 5.1-5.5 occurs in protocluster environments.\\
 
Galaxies in high-redshift overdensities often have different properties compared to those in less dense regions. For example, \cite{lemaux2022} find a highly significant positive relation between the star formation rate (SFR) and galaxy densities at $3\lesssim z \lesssim5$, with galaxies in the densest protocluster environments exhibiting average SFRs $>$0.5\,dex higher than those in the coeval field. 
At higher redshifts (z$>$4.5), \cite{Li2024} observe that galaxies in dense environments exhibit a slightly elevated SFR at a given mass compared to those in less dense regions. Additionally, they find that galaxies in high-density regions tend to have redder UV slopes, which may indicate increased dust extinction (see also \citealt{Helton2024_b}). \cite{Herard-Demanche2024} compared the properties of galaxies forming within the GOODS-N z$\sim$5.2 overdense region to those outside and found that the masses, SFRs, and UV luminosities of galaxies inside the overdensity are significantly higher than those outside.
\cite{Morishita2025} identify two overdensities of galaxies at z$\sim$5.7 along the sightline of the galaxy cluster Abell\,2744, where some members exhibit properties suggestive of accelerated maturation.
These studies show that the environment influences galaxy growth early in cosmic history even if high-z galaxies are also much more gas-rich than local ones, for the same total mass, which could also give rise to very high SFRs. \\

As part of the NIKA2 Cosmological Legacy Survey (N2CLS), we conducted very deep observations of GOODS-N at both 1.2 and 2\,mm \citep{Bing2023, berta2025,bethermin2025}. We covered 160\,arcmin$^2$ and to a depth close to the confusion noise of the Institut de Radioastronomie Millim\'etrique (IRAM) 30\,m telescope \citep{Ponthieu2025}. The large-scale overdense region at z=5.1-5.3, as observed by the FRESCO and JADES JWST programmes \citep{Herard-Demanche2024, Sun2024}, is completely covered by N2CLS. 
%(it covers $\sim$40\% of the area). 
The survey provides a unique opportunity to study in detail the obscured star formation in the over-dense environment and analyse the properties of its millimetre-bright dusty star-forming galaxy (DSFG). This paper presents the discovery of N2GN\_1\_13 at z$_{\rm spec}$=5.182, a massive DSFG belonging to the overdensity that we characterised using Northern Extended Millimetre Array (NOEMA) observations. Furthermore, by combining our analysis with JWST/FRESCO data, we revealed another N2CLS DSFG at z$_{\rm spec}$=5.201. We additionally identify three N2CLS galaxies with a photometric redshift of z$\sim$5.3, one of which is optically dark.\\

The paper is organised as follows. Section\,\ref{N2CLS} gives a summary of the N2CLS survey of GOODS-N. In Sect.\,\ref{ID13} we present the NOEMA observations of N2GN\_1\_13 and give its properties, which we then compare with those of HDF850.1 and GN10 in Sect.\,\ref{sect:compar_famous}. In Sect.\,\ref{sect:others-z5} (complemented by Appendices\,\ref{App_hyperz} and \ref{App:notes}) we present the remaining five N2CLS DSFGs identified at z$=5.1-5.3$ (two with z$_{\rm spec}$, three with z$_{\rm phot}$) and discuss their physical properties. 
In Sect.\,\ref{Sect:eff} we discuss the efficiency of the DSFGs in converting baryonic gas into stars. In Sect.\,\ref{Sect:space_densities} we report their contribution to the cosmic SFRD and present their localisation in the overdense structure. 
In Sect.\,\ref{Sect:accelerated} we compare our results with those expected from empirical, semi-analytical, and hydrodynamical models. We finally provide a summary in Sect.\,\ref{Sect:cl}.
We also examine the dust production in these high-redshift galaxies in Appendix \ref{App:dust_prod}, and present another N2CLS galaxy candidate at z$\sim$5.2 (N2GN\_1\_20) in Appendix\,\ref{App:ID20}.

Throughout the paper, we adopt a $\Lambda$ cold dark matter cosmology with $H_0=67.4$ km\;s$^{-1}$\;Mpc$^{-1}$, $\Omega_\textrm{m}=0.315,$ and $\Omega_\Lambda=0.685$ \citep{planck2020_cosmo}, a \citet{chabrier2003} stellar initial mass function (IMF), and the \citet{draine2003} frequency-dependent dust absorption coefficient, $\kappa_\nu$, re-normalised as indicated by \citet[][i.e. $\kappa_{850\mu\textrm{m}}=0.047$\,m$^2$\,kg$^{-1}$, as reference]{draine2014}.

\section{DSFGs from the NIKA2 cosmological legacy survey in GOODS-N \label{N2CLS}}

N2CLS is the deepest and largest survey ever made at 1.2 and 2\,mm, covering COSMOS and GOODS-N (Table\,1 in \citealt{Bing2023}). In GOODS-N, it reaches an average 1$\sigma$ noise level of 0.17 and 0.048\,mJy over 160 sq. arcmin. Our complete sample with a purity threshold $>$95\% comprises 65 galaxies detected with S/N$>4.2$ in a least one of the NIKA2 bands. 
Thanks to a very successful NOEMA continuum observation programme, combined with sub-millimetre array (SMA) 860\,$\mu$m \citep{Cowie2017} and deep radio observations \citep{owen2018}, all GOODS-N sources have accurate positions that allow us to search for their multi-wavelength counterparts \cite[][hereafter B25]{berta2025}.  The final sample (including multiplicities) comprises 71 individual millimetre sources. Three galaxies (two identified with NOEMA, one identified with the VLA) have no multi-wavelength counterpart and thus no redshift. We refer the reader to B25 for a complete description of the identification process and construction of multi-wavelength spectral energy distributions (SEDs). The paper also provides a full description of the data employed for the SED fitting, which have been released and are publicly accessible\footnote{\url{https://data.lam.fr/n2cls}}. We have spectroscopic redshifts for 29 sources\footnote{This includes the redshifts measured with new NOEMA and JWST observations for N2GN\_1\_13 and N2GN\_1\_61 presented in this paper.} (i.e. 41\% of the sample) and photometric redshifts for 39 sources.\\

Among the 71 individual galaxies, 16 are classified as \textit{Hubble} Space Telescope (HST)-dark (some of which are also IRAC-dark). Two HST-dark galaxies remain unidentified despite their NOEMA (N2GN\_1\_17\_b) and Very Large Array (VLA; N2GN\_1\_55) counterpart. For the other 14 galaxies (6 with spectroscopic redshifts and 8 with photometric redshifts), the average redshift is z=4.9 and the average stellar mass is M$_{\star}$= 3.9$\times$10$^{11}$\,M$_{\odot}$ (stellar masses are derived using CIGALE; see Sect\,\ref{sect:properties_ID13}). This is in agreement with the general consensus in the literature that the majority of these optically dark galaxies are heavily dust-obscured, star-forming sources at z$\sim$3-6 that dominate the high-mass end of the galaxy population at these cosmic times \cite[e.g.][]{alcade2019, wang2019, Barrufet2023, xiao2023, Gottumukkala2024}.

%%%%%%%%%%%%%%%%%%%%%%%%%%%%%%%%%%%%%%%%%%%%%%%%%
\section{N2GN\_1\_13: Identifying a new DSFG in the overdense environment of HDF850.1 \label{ID13}}

\begin{figure*}
\begin{minipage}{0.3\textwidth}
\includegraphics[width=1.05\linewidth]{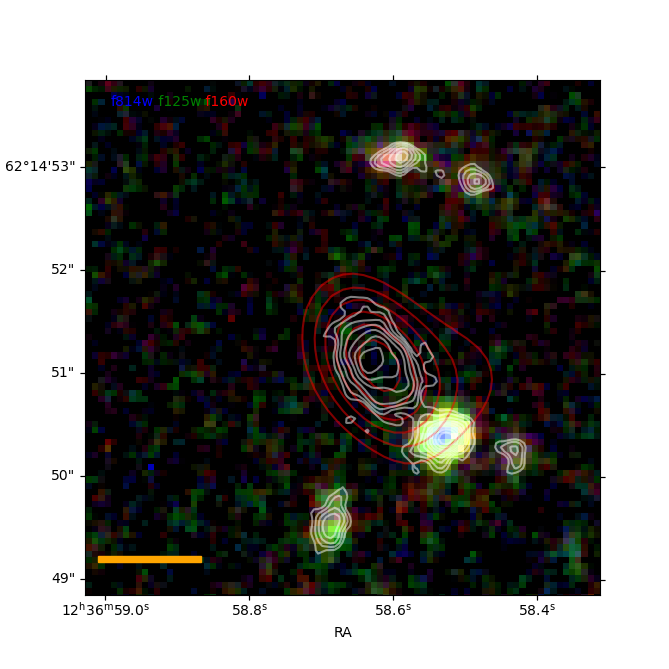}
\end{minipage}
\begin{minipage}{0.4\textwidth}
\includegraphics[width=1.\linewidth]{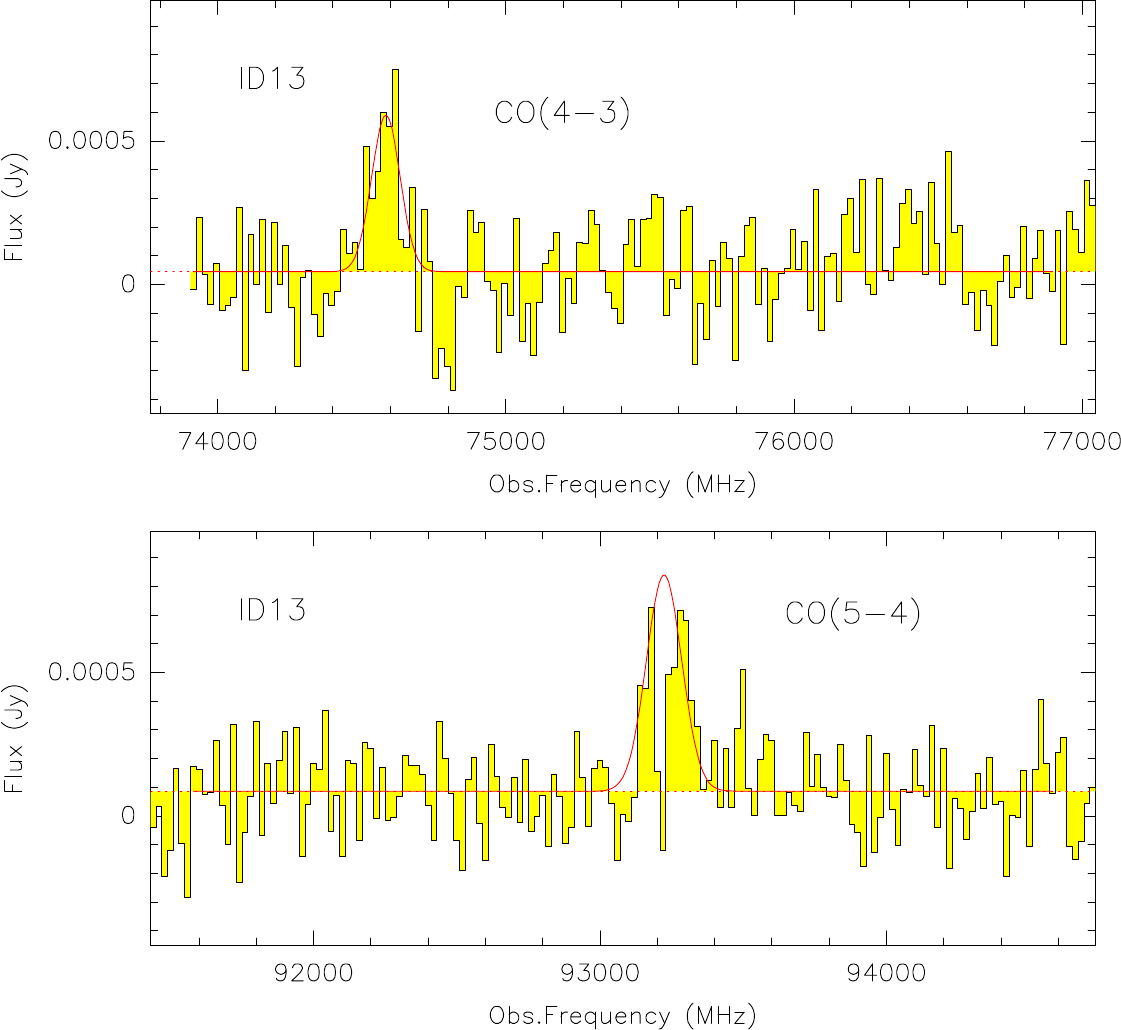}
\end{minipage}
\begin{minipage}{0.3\textwidth}
\includegraphics[width=0.9\linewidth]{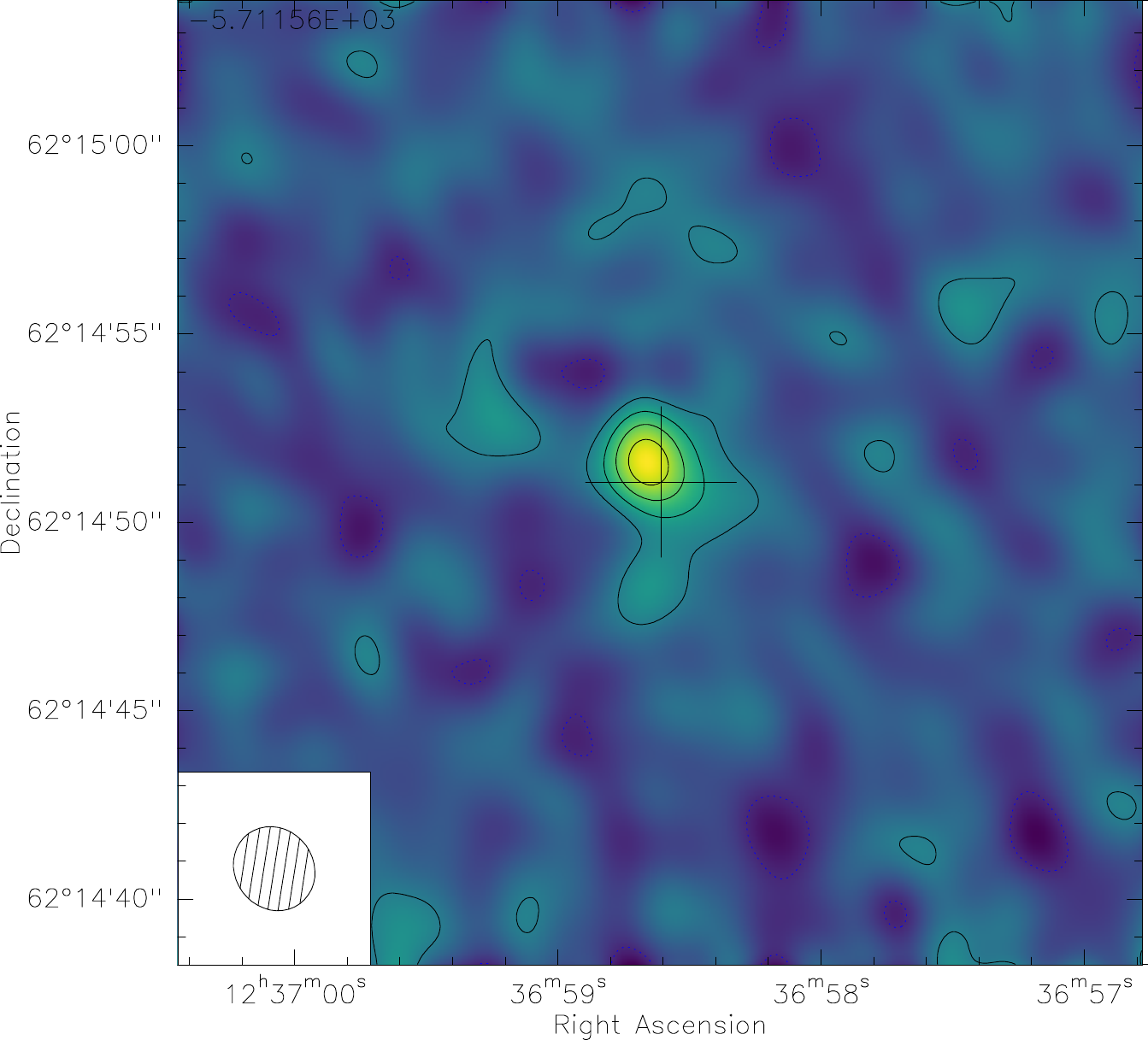}
\end{minipage}
\caption{N2GN\_1\_13. Left: False colour image produced using HST data, with F814W,
F125W, and F160W images in blue, green, and red, respectively. Overlaid in red are the NOEMA continuum contours at 2\,mm, ranging from 4$\sigma$ to 14$\sigma$ in steps of 2. The white contours are JWST F444W brightness. The scale bar in the bottom left has a length of 1". Middle:  NOEMA CO(4-3) and CO(5-4) line spectra. The red line shows the Gaussian fit to the line profile (including the continuum). Notice that the CO(5-4) line lies precisely at the interface between two basebands, resulting in the partial fragmentation of the spectral profile. Right: NOEMA 3.5\,mm continuum map. Contours are displayed with a step of 12\,$\mu$Jy/beam (corresponding to 2$\sigma$).}
\label{Fig:ID13}
\end{figure*}

We selected one of these optically dark N2CLS galaxies, N2GN\_1\_13, observed with JWST/NIRCam as part of the FRESCO programme \citep{Oesch2023} but without a confirmed redshift, as a secure target for NOEMA follow-up observations. N2GN\_1\_13 is a bright DSFG, with NIKA2 fluxes of 1.95$\pm$0.25 and 0.49$\pm$0.08\,mJy at 1.2 and 2\,mm, respectively. The accurate position of N2GN\_1\_13 has been secured thanks to our previous NOEMA programme (W21CV, P.I. L.\,Bing), which detected at 2\,mm a single counterpart with a flux of 393$\pm$47\,$\mu$Jy and a beam size of 1.61''$\times$0.93''. 
As illustrated in Fig.\,\ref{Fig:ID13} (see also Fig.\,\ref{Fig:JWST_stamps}), the galaxy is absent in the HST images
and emerges in JWST images at $\lambda\gtrsim$3.3\,$\mu$m.
The galaxy has a 6.2$\sigma$ single line detection in the NIRCam grism spectrum (Xiao, priv. comm.) but no conclusive redshift measurement. Assuming the NIRCam line is the H$_{\alpha}$ line, then z$\sim$5.196 and N2GN\_1\_13 would have the same redshift as the gigantic overdensity \citep{Herard-Demanche2024, Sun2024}.

\begin{table*}
\centering
\caption{N2GN\_1\_13. CO emission line observed frequencies, widths, fluxes, and luminosities. \label{tab:CO_ID13}}   
\begin{tabular}{c|c|c|c|c|c}
\hline \hline
Line & $\nu_{obs}$ & FWHM & S$_\textrm{line}$ &  L$^\prime_{\textrm{line}}$ & L$_{\textrm{line}}$  \\
 & [GHz] & [km\,$\rm s^{-1}$] & [Jy\,km\,$\rm s^{-1}$] & $[\rm 10^{10}\ K \, km \, s^{-1} pc^2]$ & $[\rm 10^{7}\,L_{\sun}]$  \\ \hline
CO(4-3) & 74.583 & 470$\pm$110 & 0.27$\pm$0.08 & 1.65$\pm$0.49 & 5.19$\pm$1.54 \\
CO(5-4) & 93.224 & 470$^\star$ & 0.37$\pm$0.10 &  1.45$\pm$0.39 & 8.89$\pm$2.40 \\ \hline
\end{tabular}
\tablefoot{$^\star$: The linewidth of the CO(4-3) line was adopted for the CO(5-4) line.}
\end{table*} 

\subsection{NOEMA observations \label{sect:ID13_NOEMA}}
We observed N2GN\_1\_13 with NOEMA in band-1 using the C-configuration as part of programme W23CX (PI: G.Lagache). The total on-source integration time was 6.6 hours. The receivers were tuned to cover a frequency range suitable for detecting the CO(4-3) and CO(5-4) transitions at z$\sim$5.196.

The NOEMA data were processed using the standard calibration and reduction procedures implemented in the  GILDAS\footnote{\url{https://www.iram.fr/IRAMFR/GILDAS/}} software package. This resulted in calibrated uv tables for the lower and upper sidebands, sampled with a spectral resolution of 20\,MHz. The source 1125+596 was used as the phase and amplitude calibrator. The absolute flux uncertainty is estimated to be within 10\%, while the positional accuracy is better than 0.16$''$. The final dataset achieves a synthesised beam of $2.3''$$\times$$2.1''$, with a line sensitivity of 0.7\,mJy/MHz and a continuum sensitivity of 5.7\,$\mu$Jy/beam. \\

We detect two spectral lines with S/N\,>\,6, which we unambiguously identify as CO(4-3) and CO(5-4) at z=5.182.
As expected from the CO/[CI] flux line ratio in distant galaxies and our sensitivity, the [CI](1-0) at 79.6\,GHz remains undetected with S$_\textrm{[CI]}<0.89$\, Jy\,km\,$\rm s^{-1}$ (assuming the same full width at half maximum as CO(4-3))\footnote{The 3$\sigma$ upper limit was computed as $3\sigma_v \sqrt{\mathrm{FWHM} \times \Delta v}$, where $\sigma_v$ and $\Delta v$ are the RMS noise and the velocity width of a spectral channel, respectively.}. Figure\,\ref{Fig:ID13} presents the two CO detections, fitted along with the continuum. The best-fit Gaussian parameters derived from the line profiles are listed in Table\,\ref{tab:CO_ID13}. 

To characterise the continuum emission, we combined all channels from the lower and upper sidebands, excluding the spectral windows containing line emission, and produced a double sideband continuum map (Fig.~\ref{Fig:ID13}). The map was cleaned using natural weighting with a limited number of clean components. The continuum flux was measured using two methods: integration within the 2$\sigma$ contour and direct measurement at the position of the emission peak. Both approaches yield a flux of 57$\pm$6\,$\mu$Jy at 84.6\,GHz, suggesting that the continuum source is unresolved.

\subsection{Properties of N2GN\_1\_13 \label{sect:properties_ID13}}
The total IR luminosity ($\rm L_{IR}$; integrated between 8 and 1000\,$\mu$m), SFR, stellar mass $\rm (M_{\star}$), dust mass ($\rm M_{dust}$), and dust spectral index ($\beta$) have been derived in B25 using multi-wavelength SED fitting. 
The dust mass has been obtained using \citet{draine2003} frequency-dependent dust absorption coefficient ($\kappa_\nu$), re-normalised as indicated by \citet{draine2014}, i.e. $\kappa_{850\mu\textrm{m}}=0.047$ m$^2$ kg$^{-1}$ as reference. The panchromatic optical to radio SEDs are modelled with the MAGPHYS and SED3FIT codes, in their original and high-$z$ versions \citep{dacunha2008,dacunha2015,battisti2020,berta2013}, as well as with CIGALE \citep{burgarella2005,noll2009,boquien2019}. The spectral index is obtained using a modified black-body fit to the far-IR to millimetre data. The effect of the cosmic microwave background is taken into account as described by \cite{dacunha2013}. \\

The stellar mass has been obtained using the same assumptions for CIGALE and the high-$z$ versions of MAGPHYS and SED3FIT, i.e. \citet[][BC03]{bc03} stellar populations with a \citet{chabrier2003} IMF and the \citet{charlot2000} two-components attenuation law. B25 presents the comparison of stellar masses obtained using the different codes, and shows a very good agreement, but for few outliers lying at high z. Unfortunately, the sources considered in this paper are among the outliers. For example, considering the 6 optically dark sources at z$\sim$5.2 (see Table\,\ref{Tab: z5gal}), the ratio between the high-$z$ versions of MAGPHYS or SED3FIT and CIGALE stellar mass determination ranges from 1.4 to 9.2, with an average ratio of 5.8. 
Masses up to 10$^{12}$\,M$_{\odot}$ are derived using MAGPHYS and SED3FIT (and CIGALE for N2GN\_1\_20), indicating the presence of extremely massive objects at this redshift.
Excluding any active galactic nucleus (AGN) contamination, the high flux observed in the long-wavelength NIRCam/JWST F444W filter ($\sim$0.7-1.3\,$\mu$Jy) supports by itself the very large stellar masses, i.e. M$_\star \sim$10$^{11}$-10$^{12}$\,M$_{\odot}$. Such very high stellar masses go with high extinction. 
For N2GN\_1\_13, the stellar masses from SED3FIT and CIGALE are log(M$_{\star}$)=11.68 and log(M$_{\star}$)=11.03, respectively. In the following, and to be conservative, we adopt the lower stellar mass values obtained using CIGALE. To be consistent, all the other parameters (e.g. SFR, dust mass) are taken from the CIGALE SED fitting. \\

The total molecular gas mass is obtained using $\rm M_{\rm Mol} =  \alpha_{CO} \times L^\prime_{\textrm{CO(1-0)}}\,.$
Following \citet[see also \citealt{dunne2022}]{berta2023}, we adopted a conversion factor $\alpha_{CO}$= 4.0\,M$_{\odot}$\,(K\,km\,s$^{-1}$\,pc$^2$)$^{-1}$, including the contribution of helium. We transformed the luminosity of the CO lines detected by NOEMA  into $L^\prime_{\textrm{CO(1-0)}}$
by adopting the average ratios given by \citet{Carilli2013} for sub-millimetre galaxies (SMGs), $\textrm{r}_{43/10}=0.46$, and $\textrm{r}_{54/10}=0.39$, which give $\textrm{r}_{54/43}$ (and also $\textrm{r}_{54/21}$; see Sect.\,\ref{sect:compar_famous}) consistent with our measurement. We obtain $\rm M_{\rm Mol}$= (2.3$\pm$0.6)$\times$10$^{10}$\,M$_{\odot}$ and (3.1$\pm$0.9)$\times$10$^{10}$\,M$_{\odot}$ from the CO(5-4) and CO(4-3) line luminosity, respectively. We adopted the molecular gas mass from the CO(4-3), which is detected with a higher S/N and with a measurable full width at half maximum (Table\,\ref{tab:CO_ID13}). \\

We could then derive the depletion time of the molecular gas using the molecular gas mass and the SFR:
$$\tau_\textrm{dep}=\frac{\rm{ M_\textrm{mol}}}{\textrm{SFR}}.$$
This is the time it would take for a galaxy to consume the entirety of its molecular gas reservoir, if it formed stars at the rate derived from far-IR and sub-millimetre data in a single event (not considering any mass return to the interstellar medium (ISM) from stellar winds and supernovae).
We also finally computed the gas-to-dust mass ratio (GDR) as M$_{\rm Mol}$/M$_{\rm dust}$.\\

We summarise the physical properties of N2GN\_1\_13 in Table\,\ref{Tab:properties} and show JWST/NIRCam images in Fig\,\ref{Fig:JWST_stamps}. Based on these findings, we can conclude that:

\begin{itemize}

\item N2GN\_1\_13 is a very massive galaxy (1.07$\times$10$^{11}$\,M$_{\odot}$). Using the same survey volume as FRESCO \citep{Xiao2024}, such a high mass requires a high fraction of the baryons ($\sim$30\%) to be converted into stars ($\epsilon_{\star}$; as discussed in Sect.\,\ref{Sect:eff}). This is significantly higher than what is commonly observed in lower-z galaxies \citep{Xiao2024, gentile2024}.

\item N2GN\_1\_13 has a very red continuum with $F182M-F444W=3.8$\,ABmag. As shown in B25, the optical photometry reveals a rest-frame UV excess, which is reproduced by adding an unextinguished young (ages between 10 and 100\,Myr) simple stellar population to the SED modelling. As can be seen in Fig.\,\ref{Fig:JWST_stamps}, it has an extended morphology. 
The source is very red simply because it is dusty with (i) a very high A$_{\rm v}$=3.96$\pm$0.13\,mag, indicating a large dust content and (ii) rest-frame dust emission detected by NOEMA at both 324 and 574\,$\mu$m. Furthermore, this source cannot be classified as a broad-line AGN, as the H$\rm \alpha$ line does not show any additional broadening beyond what is expected from morphological and kinematic effects. It has no X-ray counterpart and no rest-frame mid-IR data are available to verify the possible presence of an AGN-torus component contributing to its mid-IR SED.

\item The very high SFR relative to the available gas reservoir results in a short depletion time of 43\,Myr, indicative of a brief ($\le$100\,Myr) starburst phase\footnote{Adopting $\alpha_{CO}$= 1\,M$_{\odot}$\,(K\,km\,s$^{-1}$\,pc$^2$)$^{-1}$ \citep[e.g.][]{Riechers2020} would give a depletion time 4 times lower, making this source even more extreme.}. Such a low $\tau_\textrm{dep}$ points to an extreme case of star formation. Assuming no gas accretion or replenishment, and modelling the gas consumption as an exponential decay $\rm{M_{gas}(t)}=\rm{M_{gas,0}} \times e^{-t/\tau_\textrm{dep}}$, the galaxy's SFR would drop to 10\% of its initial value within approximately 100\,Myr, effectively marking the quenching of star formation.

\item With a dust mass of (7.7$\pm$0.8)$\times$10$^8$\,M$_{\rm \odot}$, and given its stellar mass, N2GN\_1\_13 lies near the maximal theoretical dust yield expected from the combined contributions of supernovae (assuming no reverse-shock destruction) and asymptotic giant branch (AGB) stars (see Fig.\,\ref{Fig:Dust_production} and  \citealt{witstok2023}). However, the short gas depletion time suggests that dust production from AGB stars, whose contribution operates on longer timescales, is likely minimal. This implies that additional dust formation mechanisms, such as rapid grain growth in the ISM, may be required to account for the observed dust mass.

\item Our observed gas-to-dust ratio leads to a galaxy metallicity consistent with solar, $Z \approx Z_\odot$ \cite[e.g.][]{li2019}.
 However, this estimate should be interpreted with caution, as our adopted values for $\alpha_{\rm CO}$ and $\kappa_\nu$ are themselves based on the assumption of solar-like metallicity.

\item With a CO(5–4)/CO(4–3) flux ratio of 1.37, N2GN\_1\_13 closely matches the ratio observed in the galaxy CGG-z4.b at $z = 4.3$ \citep{Brinch2025}, and falls within the upper range of values reported in the literature (see Fig. 5 in \citealt{Brinch2025}). This elevated ratio suggests the presence of high gas temperatures, densities, and/or pressures.

\item N2GN\_1\_13 has a very high SFR and is thus in the process of very efficient stellar mass build-up. Compared to the star-forming main sequence (SFMS), N2GN\_1\_13 lies significantly above the relation if we adopt the parametrisation of \citet{popesso2023}, with $\Delta$SFR = log(SFR/SFR$_{\rm MS}$)= 0.55$\pm$0.02 (for a scatter of 0.09\,dex). Using the parametrisation of \citet{speagle2014}, however, the offset is smaller, $\Delta$SFR = 0.29$\pm$0.04 (for a scatter of 0.2\,dex). These SFMS relations were calibrated with pre-JWST data, and it is now well established that massive galaxies at z$\sim$5 can remain undetected in HST observations. As a result, the SFMS at the high-mass end (M$_{\star}\gtrsim 10^{11}$\,M$_{\odot}$) is not necessarily well constrained. Considering the specific star-formation rate (sSFR), N2GN\_1\_13 has log(sSFR)=-8.18$\pm$0.13, which is close to the empirical limit of log(sSFR)$\lesssim$-8.05 for main-sequence (MS) galaxies reported by \citet[see also \citealt{rinaldi2025}]{caputi2017}, based on star-forming galaxies selected from optical and near-IR surveys.\\
Taken together, these results suggest that N2GN\_1\_13 is broadly consistent with, though slightly above, the SFMS, within the current uncertainties at the high-mass end.
 \end{itemize}

N2GN\_1\_13 is thus a massive, extended, and highly obscured galaxy caught in a short-lived starburst phase at z$\sim$5.2. Its high stellar mass, red optical continuum, and gas reservoir point to rapid evolution and imminent quenching.

\begin{table*}
\centering
\caption{Derived properties for z$\sim$5.2 N2CLS galaxies with CO measurements in GOODS-N. \label{Tab:properties}}   
\begin{tabular}{l|l|c|c|c}
\hline \hline
Quantity & Unit & N2GN\_1\_13 & N2GN\_1\_06 & N2GN\_1\_01 \\
& & & HDF850.1 & GN10 \\ 
& & z=5.182 & z= 5.183 & z=5.303 \\ \hline
L$_{\rm IR}$ & 10$^{12}$ L$_{\odot}$ & 6.5$\pm$0.4 & $\bar{\mu}_L^{-1} \times$(8.2$\pm$0.6) & 15.0$\pm$0.1 \\ 
SFR & M$_{\odot}$\,yr$^{-1}$ & 703$\pm$41 & $\bar{\mu}_L^{-1} \times$(899$\pm$68) &  1639$\pm$82\\  
$\beta$ &  & 1.90$\pm$0.21 & 2.56$\pm$0.36 & 2.98$\pm$0.06 \\
M$_{\rm dust}$ & $\frac{0.047}{\kappa_{850\mu\textrm{m}}}$ \,10$^{8}$ M$_{\odot}$ & 7.7$\pm$0.8 & $\bar{\mu}_L^{-1} \times$(11.9$\pm$1.5) &  16.7$\pm$0.8\\  
M$_{\star}$ & 10$^{11}$ M$_{\odot}$ &  1.1$\pm$0.1 & $\bar{\mu}_L^{-1} \times$(0.6$\pm$0.2) & 1.6$\pm$0.1 \\  \hline
M$_{\rm Mol}$ & $\frac{\rm \alpha_{\rm CO}}{4}$\,10$^{10}$ M$_{\odot}$ &  3.1$\pm$0.9 & $\bar{\mu}_L^{-1} \times$(12$\pm$5)  & 25$\pm$3 \\ 
$\tau_{\rm dep}$= M$_{\rm Mol}$/SFR & $\frac{\rm \alpha_{\rm CO}}{4}$ Myr & 43$\pm$13 & 137$\pm$54 & 155$\pm$20 \\ 
GDR= M$_{\rm Mol}$/M$_{\rm dust}$ & $\frac{\rm \alpha_{\rm CO}}{4}$ $\times$ $\frac{\kappa_{850\mu\textrm{m}}}{0.047}$ & 40$\pm$12  & 103$\pm$42 & 152$\pm$20 \\  \hline
\end{tabular}
\tablefoot{The dust and molecular gas masses are obtained using $\kappa_{850\mu\textrm{m}}=0.047$ m$^2$ kg$^{-1}$ and $\rm \alpha_{\rm CO}$= 4.0\,M$_{\odot}$\,(K\,km\,s$^{-1}$\,pc$^2$)$^{-1}$, respectively. The mean gravitational lensing magnification factor for HDF850.1 is $\bar{\mu}_L$=2.5 \citep{Sun2024}. The molecular gas mass has been computed using the lowest J CO transition available (4-3, 2-1 and 1-0 for  N2GN\_1\_13, 06 and 01, respectively; see Sect.\ref{sect:compar_famous}).}   
\end{table*}

\section{Gas masses, gas-to-dust ratios, and gas depletion times for the three brightest DSFGs in the overdense environment \label{sect:compar_famous}}

In addition to N2GN\_1\_13, two other galaxies in the overdense environment have CO measurements: GN10 and HDF850.1 \citep[see the compilation in][R20]{Riechers2020}.
We used the recipes described in Sect.\,\ref{ID13} to compute gas masses, GDR and gas depletion times for these two galaxies and use the other source properties (e.g. SFR) as derived from the SED fitting with CIGALE. The results are given in Table\,\ref{Tab:properties}.

We used the CO(2-1) line measurements for HDF850.1 and GN10 and the conversion factor from \cite{Carilli2013}, which gives M$_{\rm Mol}$ consistent with R20 if we consider $\alpha_{CO}$=1 rather than $\alpha_{CO}$=4, and assuming a mean gravitational lensing magnification factor $\bar{\mu}_L$=1.6 for HDF850.1 as in R20. If we further correct for $\kappa_{850\mu\textrm{m}}$ (which is equal to 0.057\,m$^2$ kg$^{-1}$ in R20 versus 0.047 here), we would obtain GDR=46 and 31 w.r.t. 65 and 130 in R20, and $\tau_{\rm dep}$=39 and 34\,Myr  w.r.t to 70 and 40\,Myr in R20, for GN10 and HDF850.1, respectively. 
The large difference in the GDR for HDF850.1 arises from a factor of 3.7 difference in M$_{\rm dust}$ (assuming the same $\kappa_{850\mu\textrm{m}}$), while the discrepancy in $\tau_{\rm dep}$ for GN10 is mainly due to a factor of 1.6 difference in the SFR. Notice that we have the same spectral index $\beta$ as in previous studies (R20, \citealt{walter2012}).\\

We derive a gas mass fraction (f$_{\rm gas}$ = M$_{\rm Mol}$ / (M$_{\rm Mol}$+M$_{\star}$)) of $\sim$0.6 for HDF850.1 and GN10, which is typical of gas fraction of z$\sim$5 SMGs \citep[][]{S2CLS_UDS2020}. For N2GN\_1\_13, the gas mass fraction is $\sim$3 times lower but remains consistent with certain z$\sim$5 SMGs reported in the literature \citep[e.g.][]{Zavala2022}. However, we warn that such measurements have to be taken with caution.
For instance, \cite{Sun2024} reported a gas fraction of 0.18 for HDF850.1, which is significantly lower than the value obtained in this study. This discrepancy primarily arises from the considerably lower gas mass reported in \cite{Neri2014}, attributed to the adoption of a much lower $\alpha_{CO}$ (this is also the case in \cite{Zavala2022} where $\alpha_{CO}$=1.0\,M$_{\odot}$\,(K\,km\,s$^{-1}$\,pc$^2$)$^{-1}$).\\

\citet{tacconi2020} give updated scaling relations that links M$_\star$, redshift and distance from the MS of star-forming galaxies (from \citealt{speagle2014}) to molecular gas depletion timescale ($\tau_\textrm{dep}$) and thus molecular gas mass (M$_\textrm{mol}$). We used these relations to compute $\tau_\textrm{dep}$ and M$_\textrm{mol}$ (see Table\,\ref{Tab:properties_all}) and obtain $\tau_\textrm{dep}$= 206, 142, 161\,Myr and M$_\textrm{mol}$=1.4$\times$10$^{11}$, 5.1$\times$10$^{10}$, 2.6$\times$10$^{11}$\,M$_{\odot}$ for N2GN\_1\_13, HDF850.1 (de-lensed) and GN10, respectively. While this compares quite well to the results given in Table\,\ref{Tab:properties} for HDF850.1 and GN10 ($\sim$5\% level), M$_\textrm{mol}$ (and thus $\tau_\textrm{dep}$) is overestimated by a factor of $\sim$4.5 for N2GN\_1\_13. If we used higher J transitions to get the molecular mass for HDF850.1 and GN10, we would have obtained the same discrepancy (i.e. molecular gas mass of 3.1 and 5.4$\times$10$^{10}$M$_{\odot}$ using J$_{\rm up}$=5 rather than 12 and 25$\times$10$^{10}$ M$_{\odot}$). This shows how these determinations have to be taken with caution.

\section{Bright dusty galaxies in the overdense environment \label{sect:others-z5}}

\begin{table*}
\centering
\caption{ N2CLS galaxies identified in the overdensity. 
\label{Tab: z5gal}}   
\begin{tabular}{l|l|ll|ll|l}
\hline\hline
NIKA2 ID & Alt.$^\dagger$ & RA & Dec &  z$_{\rm s,p}^\ddagger$ & z$_{\rm p}^h$ &  Comment\\
& & [hh:mm:ss] & [dd:mm:ss] & & \\
\hline
N2GN\_1\_01 & GN10, S2 & 12:36:33.44 & 62:14:08.7 &  5.303$^{\rm s}$ & &  Optically dark \\
N2GN\_1\_06 & HDF850.1 & 12:36:51.99 & 62:12:25.7 & 5.185$^{\rm s}$ & & Optically dark  \\
N2GN\_1\_13 & GN32 & 12:36:58.61 & 62:14:51.1 & 5.182$^{\rm s}$ & & Optically dark  \\
N2GN\_1\_23 & GN15, S3 & 12:36:56.54 & 62:12:07.5 & 5.179$^{\rm s}$  & & Optically dark \\ 
N2GN\_1\_61 & P-40 &  12:36:57.48 & 62:16:54.4 &  5.201$^{\rm s}$ & & Optically dark \\ \hline
N2GN\_1\_43 &  & 12:37:11.95 & 62:11:19.7 & 5.87$\pm$0.38$^{\rm p}$ & 5.31$_{-0.86}^{+1.33}$  & Optically dark\\
N2GN\_1\_26 & GN38 &  12:37:41.14 & 62:12:20.4 & 5.31$\pm$0.08$^{\rm p}$ & 4.87$_{-0.19}^{+0.21}$  &  \\
N2GN\_1\_57 & HRG14 & 12:37:31.69 & 62:16:16.5 & 5.30$\pm$0.18$^{\rm p}$ & 5.18$_{-0.20}^{+ 0.19}$ &  \\
\hline
\end{tabular}
\tablefoot{The majority are not detected in the deepest HST image, which reaches a 5$\sigma$ depth of 27.3\,mag in F160W \citep{barro2019}. These sources are labelled `Optically dark'.\\
$^\dagger$: The HDF850, GN, HRG14, S and P- nomenclatures were defined by \citet{hughes1998}, \citet{pope2005}, \cite{finkelstein2015}, \cite{Xiao2024}, \cite{greve2008}.\\
$^\ddagger$ s: spectroscopic redshift and p: photometric redshift, from B25 ;
$^h$ photometric redshift from New-Hyperz.}
\end{table*} 

\subsection{N2CLS galaxies at similar redshifts}
In addition to N2GN\_1\_01 (GN10), N2GN\_1\_06 (HDF850.1), and N2GN\_1\_13, we have five more galaxies with redshifts and position compatible with the overdense structure\footnote{N2CLS found an other dusty galaxy at z$\sim$5.3, which is farther away (see Appendix\,\ref{App:ID20}).} (see Table\,\ref{Tab: z5gal}).
Two are optically dark N2CLS galaxies with spectroscopic redshift from JWST: N2GN\_1\_23 (also known as S3 in \citep{Xiao2024} and N2GN\_1\_61.
For the three additional galaxies with previously reported photometric redshifts of $z\sim5.2$ (see B25 and references therein), independent photometric redshifts were also derived using the enhanced JWST photometric coverage (when available) and the SED-fitting code New-Hyperz. This is a modified version of the public code Hyperz originally developed by \citet{Bolzonella_2000}\footnote{\url{https://people.osupytheas.fr/~rpello/newhyperz/}}. The procedure and its validation is discussed in Appendix\,\ref{App_hyperz}. Results are shown in Table\,\ref{Tab: z5gal}. Our new photometric redshifts are within one-sigma uncertainty of the literature based $z_{phot}$, for all but N2GN\_1\_26 (2\,$\sigma$). The new photometric redshifts therefore support their possible association with the z$\sim$5.1-5.4 overdensity. While photometric redshifts provide the best constraints currently available, spectroscopic confirmation will be required to secure their membership. For example, a photometric-redshift-selected galaxy, N2GN\_1\_33, initially estimated at z$_{\mathrm phot}\sim$5.3 (B25) was later confirmed with JWST spectroscopy to lie at z=1.78, illustrating the uncertainties inherent in photometric redshift estimates. In the following, we therefore clearly distinguish results for galaxies with spectroscopic and photometric redshifts. Among the three N2CLS galaxies with z$_{\rm phot}$, one is optically dark (N1\_1\_43), while the remaining two are optically bright (N2GN\_1\_26 and N2GN\_1\_57).
Details on each of the five sources are provided in Appendix\,\ref{App:notes}.

Of the eight sources (five with $z_{\rm spec}$ and three with $z_{\rm phot}$), N2GN\_1\_01, 06, and 23 belong to the overdensity according to \cite{Sun2024}, \cite{Herard-Demanche2024}, and  \cite{Xiao2024}; N2GN\_1\_13 and N2GN\_1\_61 fall into their area of study but were not identified. 
N2GN\_1\_43 is in the JADES footprint but was not reported. N2GN\_1\_57 lies at the border of the JADES footprint and  N2GN\_1\_26 is farther away.

\begin{table*}[]
\centering
\caption{Properties of millimetre-bright DSFGs in the overdensity at z$\sim$5.2.  \label{Tab:properties_all}}   
\begin{tabular}{lllllll}
\hline\hline
NIKA2 ID & SFR & A$_{\rm V}$ & M$_{\star}$ & M$_{\rm dust}$ & M$_{\rm gas}$ &  $\tau_{\rm dep}$\\
& [M$_{\odot}$\,yr$^{-1}$] & &  [10$^{10}$ M$_{\odot}$]  &  [10$^{8}$ M$_{\odot}$]  &  [10$^{10}$ M$_{\odot}$]  & [Myr]\\
\hline
% THIRD TABLE: ADDING ERRORS ON PARAM FROM ZPHOT ERRORS (this concerns only galaxies with zphot)
N2GN\_1\_01 & 1639$\pm$82 & 4.0$\pm$0.0 & 15.8$\pm$0.8 & 16.7$\pm$0.8 & 26 & 161 \\
N2GN\_1\_06$^\dagger$  & 360$\pm$27 & 3.7$\pm$0.2 & 2.3$\pm$0.8 & 4.8$\pm$0.6 & 5.1 & 142 \\
N2GN\_1\_13 & 703$\pm$41 & 4.0$\pm$0.1 & 10.7$\pm$1.2 & 7.7$\pm$0.8 & 15 & 206 \\
N2GN\_1\_23 & 516$\pm$75 & 3.1$\pm$0.0 & 7.6$\pm$0.6 & 5.9$\pm$0.9 & 9.8 & 190 \\
N2GN\_1\_61 & 184$\pm$60 & 2.1$\pm$0.5 & 1.2$\pm$0.8 & 4.5$\pm$3.5 & 2.5 & 134 \\ \hline
N2GN\_1\_43 & 658$\pm$141 & 3.4$\pm$0.7 & 10.7$\pm$24.0 & 6.5$\pm$1.4 & 13 & 198 \\
N2GN\_1\_26 & 1962$\pm$140 & 1.3$\pm$0.0 & 9.3$\pm$1.2 & 18.3$\pm$1.4 & 23 & 118 \\
N2GN\_1\_57 & 369$\pm$49 & 2.2$\pm$0.2 & 6.0$\pm$ 1.7 & 4.2$\pm$0.9 & 8.5 & 232 \\
\hline
\end{tabular}
\tablefoot{SFR, A$_{\rm V}$, M$_{\star}$ and M$_{\rm dust}$ are derived from CIGALE fit of the multi-wavelength SEDs. The first five rows list galaxies with spectroscopic redshifts; the remaining three correspond to sources with photometric redshifts. For z$_{\rm phot}$ galaxies, errors combine fitting and redshift uncertainties (assuming $\Delta z = \pm1$).
M$_{\rm gas}$ and  $\tau_{\rm dep}$ are derived from \cite{tacconi2020} scaling relations.\\
$^\dagger$ corrected from $\bar{\mu}_L$=2.5.}\\
\end{table*}

\subsection{Source properties}
The multi-wavelength SEDs for these eight sources are shown in B25. Here, we used CIGALE to derive their properties. As in B25, we adopted the \citet{bc03} stellar models, with a \citet{chabrier2003} IMF and solar metallicity, the \citet{charlot2000} attenuation law, and the \citet{DL07} dust emission models, re-normalised as in \citet{draine2014}. A delayed and truncated star formation history was used, with e-folding time of the main stellar population of 0.1, 0.5, 1, 5, 10\,Gyr, and truncation ages of 50 or 100\,Myr. 

To assess the impact of photometric-redshift uncertainties on the derived physical parameters, we repeated the fits at z$_{\rm phot}\pm$1$\sigma$ for the three sources with photometric redshifts. The additional uncertainties were estimated from the standard deviation of the resulting physical parameters and added in quadrature to the fitting errors returned by CIGALE at z$_{\rm phot}$.

We summarise the results in Table\,\ref{Tab:properties_all}. Our sample is characterised by high V-band attenuation (A$_V$), SFR and stellar mass, with median values of A$_V$=3.26,  SFR=587\,M$_{\odot}$\,yr$^{-1}$ and log($\rm{M_\star/M_{\odot}}$)=10.93. These values are comparable to those reported for the three ultra-massive galaxies in \citet{Xiao2024}
However, they are systematically much higher than those found in the larger DSFG sample from JWST (36 galaxies) analysed in the same study. \\

\begin{figure}[]
\centering
\includegraphics[width=0.95\linewidth]{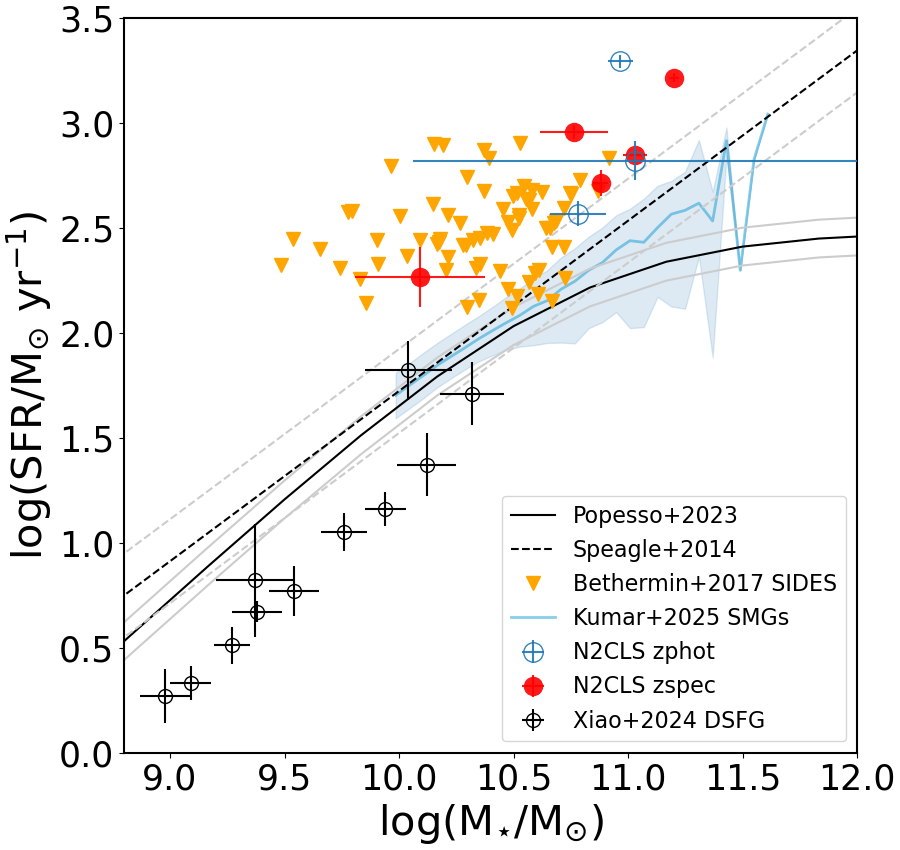}
\caption{Position of our eight N2CLS DSFGs in the SFR--M$_\star$ plane at $z = 5.2$, with sources at spectroscopic and photometric redshifts shown in red and blue, respectively. The SFMS from \citet{popesso2023} and \citet{speagle2014} is shown, with the $1\sigma$ scatter indicated by the light grey lines. For comparison, we include DSFGs from the GOODS-N field identified by \citet{Xiao2024} using JWST at $5.1 < z < 5.4$ (excluding HDF850.1, GN10, and S3, which are part of our N2CLS sample). Galaxies from the 2 deg$^2$ simulated infrared dusty extragalactic sky (SIDES) simulation \citep{Bethermin2017}, selected in a similar way as N2CLS with $S_{1.2\,\mathrm{mm}} > 0.7$\,mJy at $5.1 < z < 5.4$, are shown as downward orange triangles. The blue area indicates the location of the z=5.2 DSFGs from \cite{kumar2025}, selected based on their 850\,$\mu$m flux, using a flux threshold consistent with our N2CLS sample.}
\label{Fig:MSz5}
\end{figure}

Combining the derived SFR and M$_\star$, and the known redshift, we applied the \citet{tacconi2020} scaling relations to derive the molecular gas depletion timescale, $\tau_\textrm{dep}$, and mass, $\rm{M_\textrm{gas}}=\tau_\textrm{dep}\times\textrm{SFR}$ of these N2GN sources. 
Figure\,\ref{Fig:MSz5} locates the sources in the $\rm{M_\star}$ versus SFR plane and compares their position to the MS loci defined by \citet{speagle2014} and \citet{popesso2023}. The \citet{tacconi2020} scalings are based on the distance of the sources from the \citet{speagle2014} MS, which does not include the bending at large $\rm{M_\star}$, appearing clearly instead in \cite{popesso2023}.

The N2GN z$\sim$5.2 galaxies lie significantly above the SFMS as defined by \citet{popesso2023}, indicating a bursty mode of star formation. Although they appear closer to the SFMS when adopting the \citet{speagle2014} parametrisation, the eight galaxies still lie above the $1\sigma$ envelope of the relation. This result will need to be confirmed for such high-mass and high-SFR galaxies at such redshifts once the SFMS is better constrained with JWST data. \\

The median $\tau_\textrm{dep}$ obtained for these galaxies is of 0.18\,Gyr, with individual values spanning over the range between $\sim0.12$ and $\sim0.23$\,Gyr. These short timescales (or equivalently a very high star formation efficiency, $\epsilon=1/\tau_\textrm{dep}\sim5.6\times10^{-9}$\,yr$^{-1}$) confirm the intense ongoing starbursts activity that powers their far-IR and millimetre emission. 
Despite their short depletion timescales, the starbursts are still operating over periods significantly longer than their dynamical timescales, suggesting that the activity is sustained through multiple bursts or extended starburst phases. The fact that these galaxies reside in a confirmed overdense environment likely provides both the fuel and favourable conditions to maintain such prolonged star formation.

Given their short gas depletion timescales, these DSFGs may transition into a quiescent phase by $z \sim 4.5$, assuming a constant star formation efficiency, complete molecular gas consumption, and no external gas replenishment. This evolutionary pathway would be consistent with the emergence of massive quiescent galaxies observed in other overdense environments at $z > 4$ \citep[e.g.][]{Carnall2024, Kakimoto2024, Tanaka2024}.\\

We finally examined the rest-frame optical morphologies of our sources using JWST/NIRCam imaging (Fig.\,\ref{Fig:JWST_stamps}). Most galaxies appear extended or clumpy, with no strong evidence of point-like or compact morphologies ; only one galaxy in the sample appears marginally compact. Several galaxies show irregular or asymmetric light distributions, suggestive of ongoing or recent dynamical activity, such as minor mergers, interactions, or internal disk instabilities. These features are broadly consistent with recent JWST results for high-redshift, dust-obscured galaxies \citep[e.g.][]{McKinney2025}, which reveal a wide range of morphologies, from clumpy and irregular to disk-like systems, among massive DSFGs. 

\subsection{Impact of missing far-IR data on SFR and stellar mass estimates for our sample}
We also display in Fig.~\ref{Fig:MSz5} the locations of the DSFGs identified by \citet{Xiao2024}, selected based on their JWST colours $F182M - F444W > 1.5$\,mag, which suggest significant dust attenuation. Notably, these JWST-selected galaxies lie systematically below the SFMS. This offset may stem from underestimated SFRs and/or overestimated M$_{\star}$. In the absence of far-IR or (sub-)millimetre constraints (`No IR' case), UV-based SFRs, corrected for dust attenuation, frequently fail to capture the bulk of star formation in dusty systems. GOODS-ALMA studies \citep{Elbaz2018, xiao2023} report typical ratios of total to dust-corrected UV SFRs of the order of $\sim$8. Conversely, \citet{Pacifici2023} find that including far-IR data in SED fitting yield systematically lower SFRs compared to fits that omit such data.
To investigate this further, we performed SED fitting on a subset of our N2CLS DSFGs using only optical to near-IR photometry ($\lambda_{\rm obs} < 10\,\mu$m). Using BAGPIPES \citep{Carnall2018}, we modelled the UV to near-IR SEDs of five sources with available HST+JWST photometry (N2GN\_1\_01, 06, 13, 23 and 43). For N2GN\_1\_01 (S2) and N2GN\_1\_23 (S3), we confirm the results of \citet{Xiao2024}, with SFRs significantly underestimated. However, this is not the case for N2GN\_1\_43 where higher SFRs are recovered.
To further probe this discrepancy, we repeated the analysis using the CIGALE code. The inferred SFRs from the No IR fits differ notably from those obtained with the full UV to (sub-)millimetre coverage, with differences reaching up to a factor of $\sim$20. SFRs are overestimated in the No IR scenario, except for two sources. Importantly, the stellar masses derived in the No IR case are systematically overestimated, with offsets ranging from a factor of 1.2 up to 7.5.
These comparisons underscore that, even with the unprecedented depth of JWST, (sub-)millimetre data remain essential to accurately determining SFRs and M$_{\star}$ (and A$_V$) in dusty galaxies. However, such data require very deep observations. Indeed, none of the 25 DSFGs in the GOODS-N field reported by \citet[except for the brightest sources N2GN\_1\_01, 06, and 23]{Xiao2024} are detected at $>3\sigma$ in N2CLS, which already nearly reaches the confusion limit in GOODS-N \citep{Ponthieu2025}.

\section{Stellar baryon budget and the efficiency of star formation \label{Sect:eff}}

\begin{figure}[]
\centering
\includegraphics[width=0.95\linewidth]{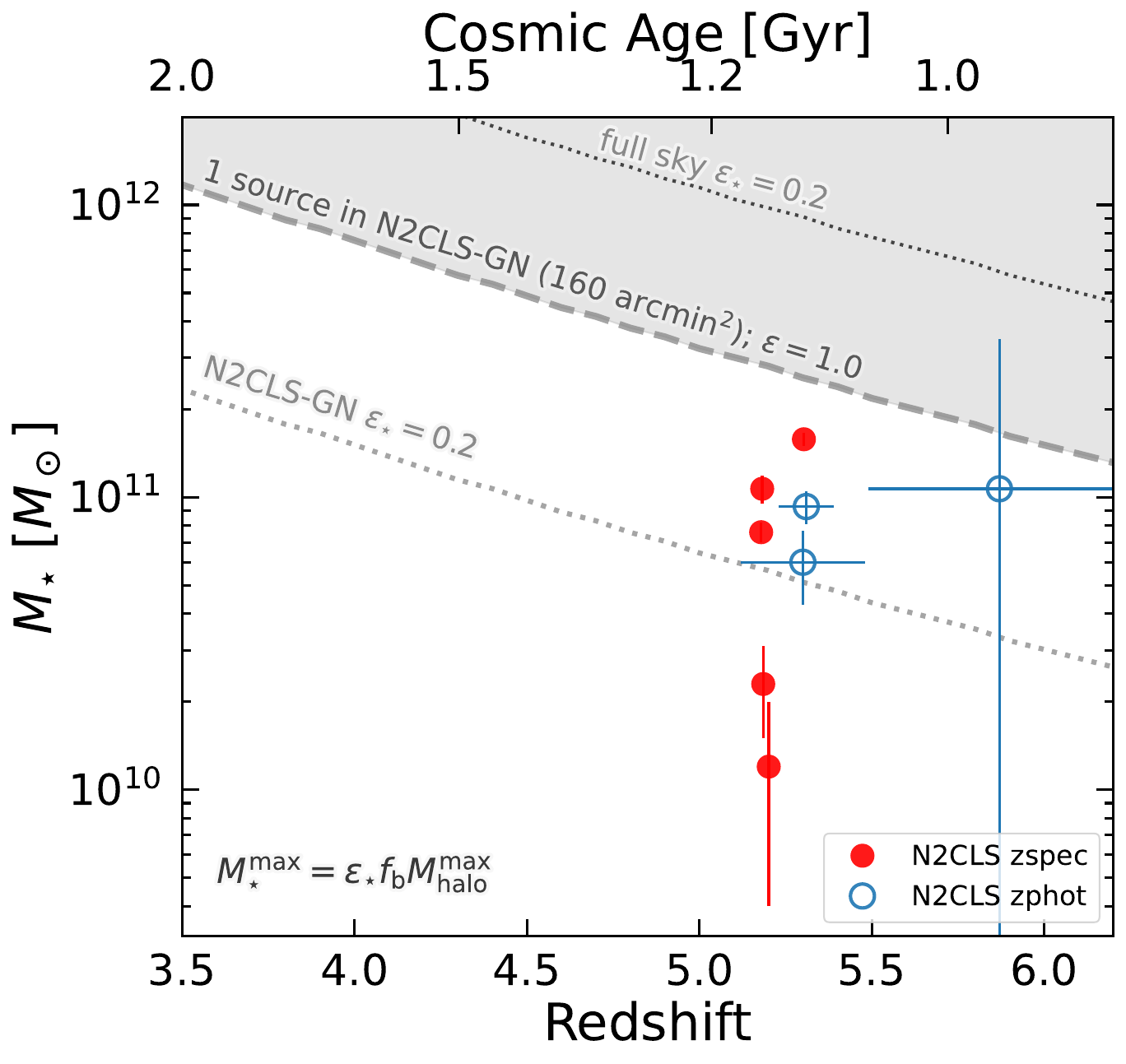}
\caption{Stellar masses of our galaxies at $z\sim5.2$ compared to expectations from the simple relation that bounds the total stellar content of a dark matter halo to $M_{\star} \le \epsilon_{\star}\,f_b\,M_{halo}$, where $\epsilon_{\star}$ (which should be $\leq 1$) is the efficiency with which baryons are converted into stars and $f_b$ is the cosmic baryonic fraction. We indicate the maximum stellar mass ($M_{\star}^{\rm max}$) calculated from the maximum halo mass given our survey volume (160\,arcmin$^2$, z=5.2$\pm$0.5) for $\epsilon_{\star}$=0.2 (dotted line) and $\epsilon_{\star}$=1 (dashed grey line). We also show $M_{\star}^{\rm max}$ for $\epsilon_{\star}$=0.2 using the full sky area (dotted line).}
\label{Fig:SFE}
\end{figure}

Recent observations have revealed a substantial population of massive galaxies in the early Universe, implying star-formation efficiencies that exceed the highest values inferred from lower-redshift studies. This may indicate a distinct and more intense mode of star formation during the earliest phases of galaxy evolution, which remains challenging for current models to reproduce \citep[][and references therein]{shen2025}.

Studies leveraging data from the JWST have identified galaxies at redshifts $z\sim5$ to $z\sim9$ with stellar masses approaching 10$^{11}$\,M$_{\odot}$. For instance, utilising JWST/Mid-infrared instrument (MIRI) data, \cite{Wang2025} found high densities of massive galaxies at $z>5$, implying elevated stellar baryon efficiencies $\epsilon_\star$. Specifically, they report that the most massive galaxies at $z\sim8$ require $\epsilon_\star$$\sim0.3$, compared to $\epsilon_\star$$\sim0.14$, for typical low-redshift galaxies.
Similarly, \cite{Xiao2024} examined 36 massive, dust-obscured galaxies with spectroscopic redshifts between $z\sim 5$ and $z\sim 9$ from the JWST FRESCO survey. They found that three ultra-massive galaxies (log(M$_\star$/M$_{\odot})\gtrsim$11.0, among them S2$\equiv$N2GN\_1\_01 and S3$\equiv$N2GN\_1\_23) require an exceptional $\epsilon_\star$ of approximately 50\%, which is two to three times higher than the efficiencies observed in galaxies at later epochs. 
Further, \cite{gentile2024} reported two massive, dusty starburst galaxies discovered in the COSMOS-Web survey, with star-formation efficiencies of $\epsilon_\star\sim$0.25 at $z_{\rm spec}$=5.051 and $\epsilon_\star\sim$0.5 at $z_{\rm phot}$=6.7$\pm$0.3, respectively.
In numerical simulations, $\epsilon_\star$ is found to increase significantly with both stellar mass and redshift, indicating that, at fixed mass, halos at (very) high redshifts convert gas into stars more efficiently than their lower-redshift counterparts \citep[e.g.][]{Ceverino2024}. 
All these results imply that certain populations of high-redshift galaxies experience very efficient star formation at early times.\\

Building upon these studies, we computed the stellar baryon conversion efficiency ($\epsilon_\star$) for our sample of galaxies at $z\sim5.2$, following the methodology adopted in \cite{Boylan-Kolchin2023} and \cite{Xiao2024}.
Using the cumulative halo mass function from \cite{tinker2008}, we computed the comoving number density of halos above a given halo mass threshold, at redshift $z$. The maximum stellar mass ($M_{\star}^{\rm max}$) that is statistically allowed within an observed volume is given by $M_{\star}^{\rm max} = \epsilon_{\star}\,f_b\,M_{halo}^{\rm max}$, where $\epsilon_{\star} \leq 1$ represents the (integrated) efficiency with which baryons are converted into stars and $f_b\equiv\Omega_b/\Omega_m$=0.156 is the cosmic baryon fraction from \cite{planck2020_cosmo}. We show in Fig.\,\ref{Fig:SFE} our 8 galaxies in the stellar mass versus redshift plane. The lines correspond to different number densities: 
2.2$\times$10$^{-6}$\,Mpc$^{-3}$ (i.e. one source in the cosmic volume explored by N2CLS, assuming z=5.2$\pm$0.5 ; with $\epsilon_{\star}=0.2$ and  $\epsilon_{\star}=1$ displayed as blue-dotted and grey-dashed lines, respectively) and  2.4$\times$10$^{-12}$\,Mpc$^{-3}$ (i.e. a single galaxy in the whole sky; with $\epsilon_{\star}=0.2$ displayed as a dotted grey line).\\

Our analysis reveals notably high $\epsilon_\star$ values ($\epsilon_\star >$0.2 for all but N2GN\_1\_06 and 61), indicating that a significant fraction of baryonic mass in these galaxies has been converted into stars. The elevated values of $\epsilon_\star$ observed in our sample suggest that these galaxies are undergoing unusually efficient star formation at $z = 5.2$. Given their probable high metallicities and host halo masses exceeding $10^{12}$\,M$_\odot$, it is unlikely that cold gas inflows are the primary driver of this efficiency.  
Instead, the observed efficiency may reflect merger-driven inflows that channel gas into central regions and ignite intense starbursts, or compaction events that funnel enriched gas into central regions, triggering intense, short-lived starbursts.

\section{Spatial distribution and abundance of N2CLS DSFGs in the GOODS-N overdensity\label{Sect:space_densities}}

\subsection{Contribution to the star formation rate density at $z\sim5.2$ \label{Sect:space_densities_SFRD}}

Recent studies have shown that a significant fraction of the cosmic SFRD at high redshifts can be driven by rare, massive galaxies with high star formation efficiencies. For instance, \citet{Xiao2024} found that at $z\sim5.8$, ultra-massive galaxies with $\epsilon_{\star} > 0.2$ can account for up to $17^{+8}_{-3}$\,\% of the total SFRD (from \citealt{Madau2014}). These findings highlight the critical role of massive, DSFGs in early cosmic star formation, motivating a closer look at their contribution in overdense environments such as GOODS-N.\\

Based on the five N2CLS galaxies with spectroscopic redshifts, we derive a dust-obscured cosmic SFRD\footnote{Assuming an area of 160\,arcmin$^2$ and $5.1<z<5.4$ to compute the volume.}  of $\psi_{\rm spec}^{\rm IR}=(2.54\pm0.10) \times 10^{-2}$ M$_\odot$\,yr$^{-1}$\,Mpc$^{-3}$. Including the additional contribution from sources with photometric redshifts raises the SFRD to $\psi_{\rm spec+phot}^{\rm IR}=(4.77\pm0.18) \times 10^{-2}$ M$_\odot$\,yr$^{-1}$\,Mpc$^{-3}$. The quoted uncertainties combine Poisson errors and uncertainties on individual SFR estimates in quadrature. \\

\begin{figure}
\centering
\includegraphics[width=0.99\linewidth]{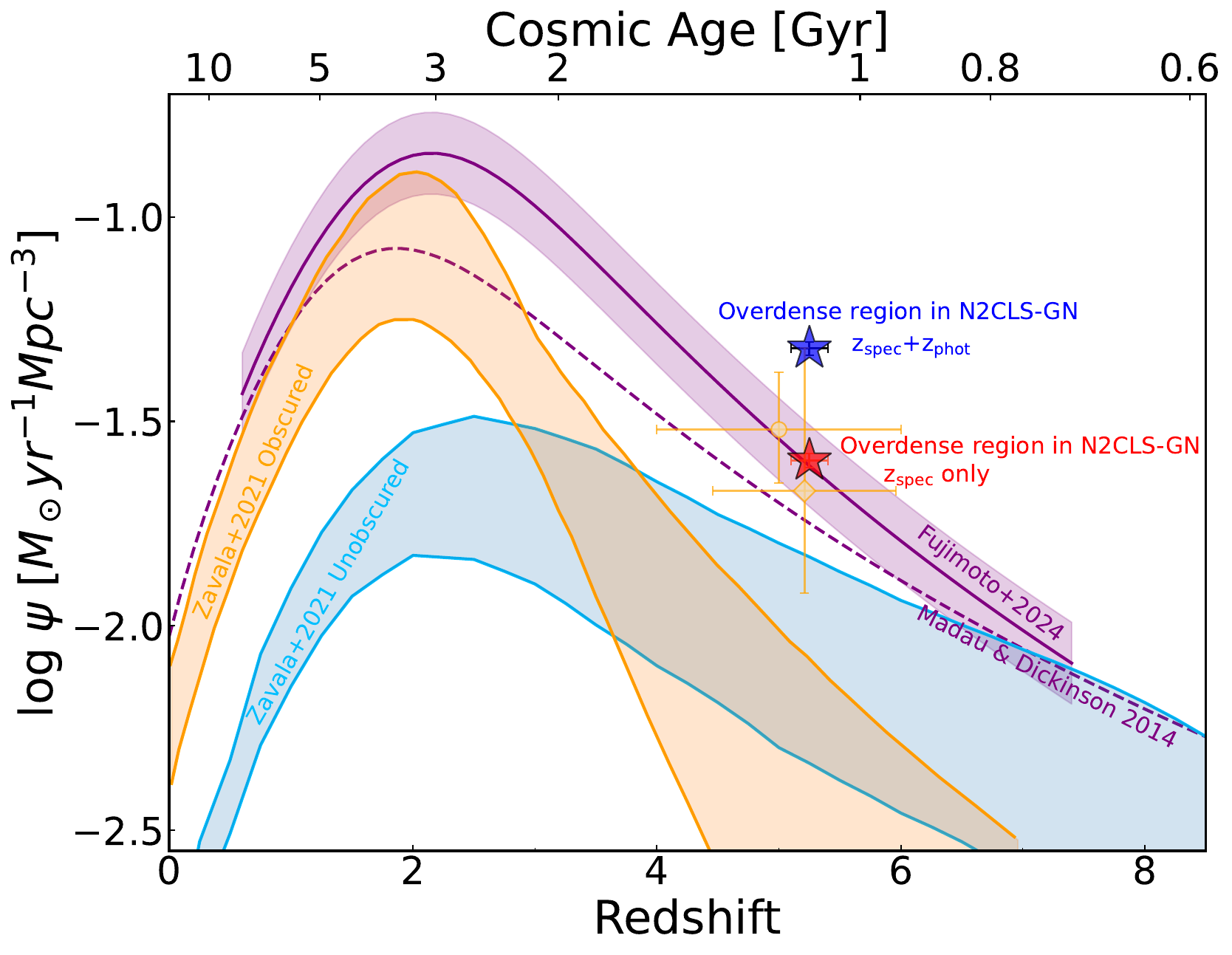}
\caption{Contribution of the dusty galaxies in the GOODS-N overdensity to the cosmic SFRD, $\psi$, at z$\sim$5.2. The red star represents the SFRD derived from spectroscopically confirmed galaxies, while the blue star also includes the contribution from galaxies with photometric redshifts. The purple curves show the total SFRD as a function of redshift: the dashed line corresponds to the \citet{Madau2014} result, obtained by correcting the UV SFRD for dust attenuation, and the solid line to \citet{fujimoto2024}, obtained by summing the obscured (IR) and unobscured (UV) contributions. At z$\sim$5, the dust-obscured SFRD estimates from \citet{Traina2024} and \citet{sun2025} are shown as an orange diamond and circle, respectively. The orange and blue shaded areas represent the obscured and unobscured SFRD, respectively, from \cite{Zavala2021}.}
\label{Fig:SFRD}
\end{figure}

Figure~\ref{Fig:SFRD} illustrates the contribution of the N2CLS galaxies in the $z \sim 5.2$ overdensity to the cosmic SFRD. For reference, we compared our measurements to the widely used compilation of \citet[MD14]{Madau2014}. The SFRD derived from spectroscopically confirmed galaxies, $\psi_{\rm spec}^{\rm IR}$, is $1.39 \pm 0.06$ times higher than the total SFRD reported by MD14 at similar redshift. When including galaxies with photometric redshifts, the excess increases to a factor of $2.62 \pm 0.10$. However, it is now well established that the MD14 compilation underestimates the SFRD at z$\sim$4-6. ALMA studies \citep[e.g.][]{fujimoto2024, Traina2024, sun2025, liu2025} and JWST H$\alpha$ luminosity function measurements \citep{Covelo-Paz2025, Fu2025} both point to a systematically higher cosmic SFRD at these epochs. Using the updated estimate from \citet{fujimoto2024}, the contrast between our overdensity and the global average decreases to about unity for $\psi_{\rm spec}^{\rm IR}$ and to $\sim$2 for $\psi_{\rm spec+phot}^{\rm IR}$. Restricting the comparison to the obscured component, our $\psi_{\rm spec}^{\rm IR}$ is already consistent with the levels reported by \citet{Traina2024} and \citet{sun2025}, and rises substantially above them once photometric redshift members are included. Relative to the dust-obscured SFRD from \citet{Zavala2021} at similar redshifts, $\psi_{\rm spec}^{\rm IR}$ alone is more than five times higher.
If all galaxies with photometric redshifts are spectroscopically confirmed, the elevated SFRD in our sample will reflect the fact that these sources reside in a confirmed overdense environment, where star formation activity is strongly boosted compared to average field regions at the same epoch.

Beyond our specific measurement, current research suggests that massive galaxy overdensities at high redshifts may be relatively common \citep[e.g.][]{Kashino2023,Wang2023,helton2024,Brinch2024} and may thus contribute significantly to the SFRD. In the GOODS-N field, \citet{Lin2025} estimate the fraction of the cosmic SFRD hosted within the protocluster environment at z$\sim$5.2, based on the number density of H$\alpha$ emitters. They find that approximately 55\% of the SFRD at this redshift is associated with the protocluster, somewhat higher than the $\sim$30\% predicted by simulations \citep{Chiang2017} but in good agreement with recent results based on the FLAMINGO simulation suite \citep{Lim2024}. Within the same volume considered in \citet[i.e. 62\,arcmin$^2$ and $5.15<z<5.29$]{Lin2025}, our four DSFGs yield a SFR density of $\log(\rho_{\rm SFR}$\,[M$_\odot$\,yr$^{-1}$\,Mpc$^{-3}$]) = $-1.39^{+0.026}_{-0.027}$. This value is 1.6 times higher than the estimate based on the $>$150 H$\alpha$ emitters alone, suggesting that a substantial fraction of star formation activity is missed even in deep JWST surveys of highly overdense environments.

\subsection{Environmental localization of N2CLS DSFGs in the GOODS-N overdensity}
Recent JWST observations from the JADES and FRESCO programmes have revealed that the overdensity surrounding HDF850.1 represents one of the richest environments identified to date at $z > 5$, with approximately 100 galaxies at $z$=5.17–5.20 distributed across 10 substructures within a $\sim$(15\,cMpc)$^3$ volume \citep{Herard-Demanche2024, Sun2024}. This region is particularly extreme, as $\sim$28\% of all $z = 5.0$–6.0 H$\alpha$ emitters detected across the GOODS-N field are concentrated within a narrow redshift slice of $\Delta z \sim 0.03$. The spatial extent of this protocluster is consistent with the typical size of such structures predicted by \citet{Chiang2017} using a set of N-body simulations and semi-analytic models. In addition, \citet{Calvi2023} identified the galaxy GN10 (N2GN\_1\_01) as residing in an overdense region based on photometric redshifts, a result that has now been confirmed spectroscopically by \citet{Sun2024}.

\begin{figure*}[]
\centering
\includegraphics[width=0.98\linewidth]{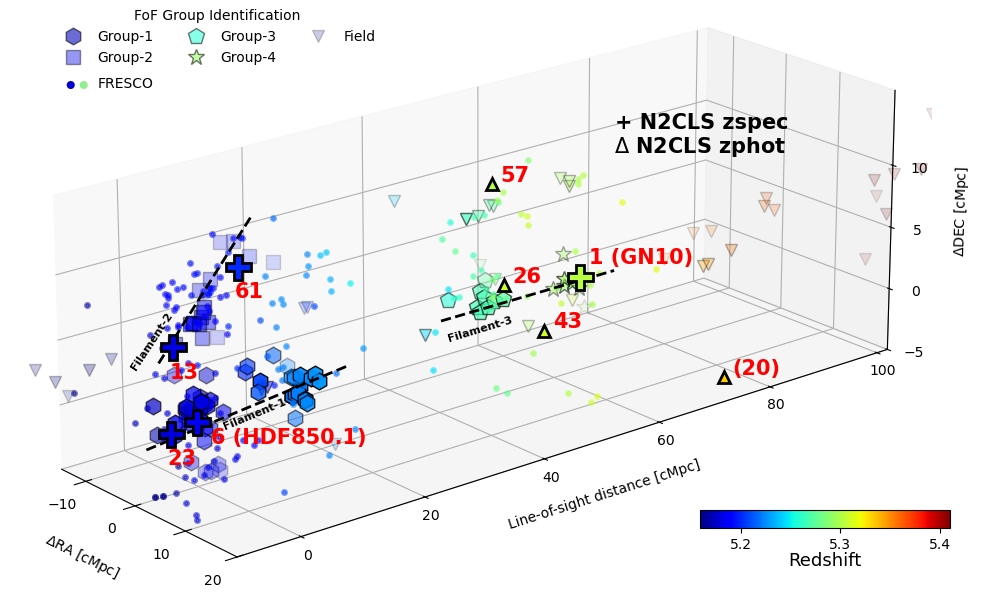}
\caption{Large-scale structure of the overdense environment in the GOODS-N field at $z = 5.17-5.40$. Galaxy positions are shown relative to HDF850.1, in units of comoving Mpc. All galaxies are colour-coded by redshift. The eight N2CLS galaxies are indicated by crosses (for spectroscopic redshifts, $z_{\mathrm{spec}}$) and upward pointing triangles (for photometric redshifts, $z_{\mathrm{phot}}$); their NIKA2 identifiers are labelled in red, with source 20 discussed in Appendix\,\ref{App:ID20}. Galaxy groups and filaments are taken from the JADES survey \citep{Sun2024}, with galaxies in less dense regions shown as transparent symbols. Additional galaxies, shown as dots, are drawn from the FRESCO survey \citep{Herard-Demanche2024}.}
\label{Fig:3D}
\end{figure*}

We show the locations of the N2CLS galaxies within the overdensity in Fig.\,\ref{Fig:3D}. Cosmological simulations suggest that at $z \gtrsim 5$, approximately 90\% of baryonic and dark matter resides in unvirialised, filamentary large-scale structures \citep[e.g.][]{Haider2016}. \citet{Sun2024} interpret the 3D spatial distribution of group member galaxies as being consistent with three distinct filamentary structures. For simplicity, we compared the positions of our galaxies relative to the filaments identified in \citet{Sun2024}, rather than to the 18 substructures detailed in \citet{Herard-Demanche2024}.
The $z = 5.17$–5.20 overdensity, traced by filament-1 and filament-2, spans the full extent of the GOODS-North FRESCO field (62\,arcmin$^2$). Within this region lie four N2CLS DSFGs with spectroscopic redshifts: N2GN\_1\_06 and N2GN\_1\_23 are located in filament-1, while the newly identified N2GN\_1\_13 and N2GN\_1\_61 fall within filament-2.
In filament-3, N2GN\_1\_26 (photometric redshift) is found along with GN10, and N2GN\_1\_43 lies nearby. N2GN\_1\_57 may trace a distinct, possibly new, structure. We note that some previously confirmed spectroscopic members of the $z \sim 5.2$ protocluster \citep{walter2012,calvi2021} fall outside the JADES and FRESCO joint footprint, implying that the overdensity exhibits more complex substructures than currently captured.

We also show the location of N2GN\_1\_20, another N2CLS galaxy at $z_\textrm{phot}=5.33$ (see Appendix\,\ref{App:ID20}), which lies farther from the core overdensity. 
Interestingly, \citet{Jiang2018} report a giant protocluster at $z \sim 5.7$ occupying a volume of $\sim 35^3$\,cMpc$^3$, embedded within a larger overdense region spanning $53 \times 41$\,cMpc$^2$ and extending 34\,cMpc along the line of sight. Further investigation is needed to determine whether the overdensity could extend towards the region traced by N2GN\_1\_20.
Finally, we note that the spatial distribution of N2CLS galaxies at z$\sim$5.2, when viewed in two dimensions (RA, Dec), does not exhibit a recognisable filamentary pattern, unlike the tentative identification of a filament with NIKA2 in the GJ526 field \citep{Lestrade2022}.

\section{Accelerated galaxy evolution in the $z\sim5.2$ overdense structure \label{Sect:accelerated}}

\citet{Chiang2017} outlined three evolutionary stages of cluster formation: an early inside-out growth phase from $z = 10$ to $z \sim 5$, an extended period of vigorous star formation from $z \sim 5$ to $z \sim 1.5$, and a final stage of rapid mass assembly and quenching from $z \sim 1.5$ to $z = 0$. Our results are consistent with the second phase of this sequence, during which intense star formation is expected to occur across 10–20\,cMpc-scale structures, contributing significantly to the stellar mass of present-day clusters. Interestingly, \citet{Lim2024}, using the FLAMINGO simulation suite, suggest that the suppression of star formation in protocluster environments already began by $z \approx 5$, indicating that this epoch could represent a transitional phase in cluster evolution.\\

Recent results from the FOREVER22 project \citep{FOREVER22} indicate that the elevated star formation activity observed in protocluster galaxies at z$\ge$6, compared to field galaxies, is primarily driven by a higher abundance of massive galaxies rather than enhanced sSFRs. This trend is likely a consequence of the denser environments in protoclusters, where increased galaxy interactions and sustained gas accretion promote the rapid assembly of massive systems. These massive galaxies tend to exhibit stronger dust attenuation, resulting in redder UV continuum slopes. This is consistent with the findings of \citet{Helton2024_b}, who reported that the UV-brightest and UV-reddest galaxies are preferentially located in denser environments with a higher number of neighbouring galaxies. Together, these observations provide compelling evidence that galaxy evolution is accelerated within high-redshift overdense structures and may help explain the record number of bright DSFGs we identified in this particular structure.\\

In the FoREVER22 project, protocluster galaxies show a combination of stellar masses and SFRs that are somewhat lower, but still comparable to those of the DSFGs identified in our study. This places our DSFGs at the upper end of the FoREVER22 population in terms of total star formation activity. In contrast, the FLAMINGO simulations from \citet{Lim2024} produce protocluster galaxies with high SFRs that are consistent with the SFRs measured for our sample. This is not the case for the general DSFG population modelled in FLAMINGO \citep{kumar2025}, selected based on their 850\,$\mu$m flux (S$_{850,\rm \mu m} > 0.3$\,mJy, corresponding to the N2CLS flux cut S$_{1.2,\rm mm} > 0.7$\,mJy at $z = 5.2$), which predominantly lies on the SFMS (see Fig.~\ref{Fig:MSz5}).
Furthermore, \citet{Chiang2017} estimate that the total SFR in protoclusters at $z \sim 5$ is typically below 1000\,M$_\odot$\,yr$^{-1}$, which is about four times lower than the combined SFR of our five spectroscopically confirmed DSFGs. The unusually high level of star formation highlights the exceptional nature of this structure. 

In the context of the SIDES simulation \citep{Bethermin2017, Gkogkou2023}, where no environmental effects are considered, we can use this dataset to address whether the enhanced dusty star formation observed here is driven by clustering or by the presence of large gas masses. To explore this, we selected SIDES galaxies in the 2-deg$^2$ simulation in the redshift range $5.1<z<5.4$ and with a NIKA2 1.2\,mm flux greater than 0.7\,mJy, which mimics the selection criteria for our sample (B25). The SFR--M$_{\star}$ relation for a NIKA2-like population is shown in Fig.\,\ref{Fig:MSz5}. With the exception of N2GN\_1\_61, SIDES galaxies generally exhibit lower SFR and M$_{\star}$.
In SIDES, the SFR is capped at a hard limit of 1000\,M$_\odot$/yr, and each galaxy's SFR is redrawn until it falls below this threshold.
This is explaining the lack of SFR$>$1000\,M$_{\odot}$/yr galaxies in SIDES. 
However, we can also see that SIDES does not reach high stellar masses. The most massive galaxy in SIDES has a stellar mass of log(M$_{\star}$/M$_{\odot}$) = 10.92. When considering the larger 117-deg$^2$ SIDES-UCHUU simulation \citep{Gkogkou2023}, the most massive galaxy reaches log(M$_{\star}$/M$_{\odot}$) = 11.29. 
SIDES employs abundance matching to obtain the stellar mass, which connects halo to galaxy stellar mass via mass functions. This approach may  suppress the presence of high-stellar mass objects if they are not represented in the observed stellar mass function used to build the SIDES simulation.
Indeed, using the stellar mass function from SIDES, we expect 0.4 galaxies with log(M$_{\star}$/M$_{\odot})>$10.7 at $5.1<z<5.4$ in the GOODS-N area (160\,arcmin$^2$), whereas we observe six (or seven if N2GN\_1\_20 is included). 

Given the non-uniform distribution of sources, we investigated the impact of field-to-field variance by dividing the SIDES-UCHUU simulation into 2,562 independent sub-fields, each covering 160\,arcmin$^2$. Among these, 13 sub-fields contain 4 galaxies above the stellar mass threshold, a small fraction but consistent with our spectroscopic ($z_{\rm spec}$) sample.
To evaluate the effect of stellar mass uncertainties, we added a random scatter of 0.5\,dex to the SIDES stellar masses and repeated the sub-field analysis. With this added uncertainty, the average stellar mass of the most massive galaxy in the redshift range $5.1 < z < 5.4$ across all sub-fields is $\log(\rm{M_{\star}/M_{\odot}}) = 11.28$, consistent with the most massive spectroscopically confirmed galaxy in our sample.
Furthermore, we now identify 25 sub-fields (1\% of the total) that contain 8 galaxies with $\log({\rm M_{\star}/M_{\odot}}) > 10.7$ within $5.1 < z < 5.4$. In conclusion, while such overdensities are rare, their existence is plausible when considering both the strong clustering of massive galaxies at $z > 5$ and the significant uncertainties in stellar mass estimates.\\

Based on their stellar masses, the five $z_{\rm spec}$ galaxies are each expected to reside in massive dark matter halos, with a combined halo mass of ${\rm M_{\rm halo}} \sim 5 \times 10^{13}$\,M$_{\odot}$, as inferred from the empirical $\rm{M_{\star}}$–$\rm{M_{\rm halo}}$ relation at $z = 5$ \citep{Behroozi2019}. While it is uncertain whether all these galaxies will eventually merge into a single structure by $z=0$, such an outcome would imply a descendant halo more massive than that of a typical Coma-like cluster progenitor at the same redshift, which simulations suggest is closer to $\sim 10^{13}$\,M$_\odot$ \citep{Chiang2013}. 
This highlights the exceptional mass scale of the environment traced by these galaxies. However, it is important to note that total mass estimates for protoclusters are sensitive to the choice of aperture and redshift interval used for their definition, influencing such comparisons across studies \citep{Lim2024}.

\section{Conclusion \label{Sect:cl}}
Using 1.2\,mm observations of GOODS-N taken with NIKA2, we identified eight N2CLS galaxies at $z\sim5.2$ (five with z$_{\rm spec}$ and three with z$_{\rm phot}$), many of which are extremely faint or even undetected in deep HST and some JWST imaging. These galaxies have remarkably high SFRs, substantial dust content, and significant stellar masses, placing them among the most extreme star-forming systems known at this epoch. Six are securely associated with the exceptional overdensity reported in \citet{Sun2024} and \citet{Herard-Demanche2024}. 

We first identified a new DSFG in the overdense environment of HDF850.1 through NOEMA observations, which detect CO(4–3) and CO(5–4) lines at $z=5.182$. In addition, another N2CLS galaxy shows a 7$\sigma$ emission line in FRESCO data, consistent with H$\alpha$ at z$_{\rm spec}=5.201$. Combined with previously known galaxies (HDF850.1, GN10, and S3), this brings the number of spectroscopically confirmed DSFGs within the overdense structure to five. We further identified three DSFGs with photometric redshifts consistent with the overdensity.

These galaxies exhibit exceptionally high SFRs (ranging from $\sim$200 to 2000\,M$_{\odot}$,yr$^{-1}$) and high stellar masses (log(M$_{\star}$/M$_{\odot}$) = 10–11), placing most of them significantly above the SFMS at $z\sim5.2$, especially under the \cite{popesso2023} parametrisation. Their properties are comparable to those of the ultra-massive galaxies analysed by \cite{Xiao2024}, suggesting an early and efficient buildup of both stellar and dust mass. They show short gas depletion timescales (with a median of 0.18\,Gyr) and remarkably high stellar baryon conversion efficiencies ($\epsilon_{\star}>$20\%, compared to $\lesssim$10\% locally). 
The observed dust masses further support the scenario of efficient dust formation. Due to their short gas depletion timescales, these DSFGs could evolve into quiescent galaxies by $z \sim 4.5$, assuming no new gas inflow. This scenario aligns with the early formation of massive quiescent galaxies seen in similar dense environments.

Their contribution to the cosmic SFRD is significant: even considering that it represents only the obscured component, the five spectroscopically confirmed galaxies exceed the total SFRD from \citet{Madau2014} by a factor of 1.4, which rises to 2.6 when including all eight sources. Compared to more recent SFRD estimates at z$\sim$5 \citep[e.g.][]{fujimoto2024}, the excess is reduced to roughly unity for the spectroscopic sample and by about a factor of 2 when including photometric redshift members. These few dusty galaxies dominate the star formation budget within the overdensity, contributing more than the much larger population of H$\alpha$ emitters.

Our DSFGs either lie at the upper extremes of stellar mass, dust mass, and SFR or are entirely absent from current simulations and models. This highlights the difficulty existing models face in reproducing such extreme systems. Their properties may be driven by efficient gas inflows along cosmic filaments in protocluster environments fuelling intense star formation despite short depletion timescales \cite[e.g.][]{Narayanan2015, Umehata2019}.

We examined the spatial distribution of N2CLS galaxies within the overdensity, comparing their positions to the three filamentary structures identified by \cite{Sun2024}. Four spectroscopically confirmed DSFGs lie along filament-1 and filament-2, two others align with filament-3, and two may trace new separate structures. We also identified an extremely massive ($\sim$3$\times$10$^{12}$\,M$_\odot$) galaxy candidate at z=5.33 that lies farther away (see Appendix\,\ref{App:ID20}). The protocluster may therefore be more extended than previously mapped, as some galaxies lie beyond current survey footprints. Confirming this will require spectroscopic redshifts for sources currently identified via photometric estimates.
If all these galaxies merged by $z=0$, they would form a halo significantly more massive than a typical Coma-like cluster progenitor at that epoch, emphasising the exceptional mass scale of the environment traced by these galaxies.

The presence of such massive, dusty, and intensely star-forming galaxies at $z \sim 5.2$ has significant implications for our understanding of galaxy formation and evolution. It suggests that the processes leading to rapid stellar and dust mass assembly were already in place within the first billion years of cosmic history in overdense environments. Future observations, particularly with facilities capable of probing the ISM properties and kinematics of these galaxies, will be crucial in unravelling the mechanisms driving their evolution.

\begin{acknowledgements}
We acknowledge financial support from the Programme National de Cosmologie and Galaxies (PNCG) funded by CNRS/INSU-IN2P3-INP, CEA and CNES, France, and from the European Research Council (ERC) under the European Union's Horizon 2020 research and innovation program (project CONCERTO, grant agreement No 788212).

This work is based on observations carried out under project numbers 192-16 with the IRAM 30-m telescope, and projects  W21CV, W23CX, and S24CF with NOEMA. 
IRAM is supported by INSU/CNRS (France), MPG (Germany) and IGN (Spain). 

This work is also based on observations made with the NASA/ESA/CSA \textit{James Webb} Space Telescope. The data were obtained from the Mikulski Archive for Space Telescopes at the Space Telescope Science Institute, which is operated by the Association of Universities for Research in Astronomy, Inc., under NASA contract NAS 5-03127 for JWST. The observations are associated with programs \#1181, 1895 and 3577. Some of the data products used here were retrieved from the Dawn JWST Archive (DJA). DJA is an initiative of the Cosmic Dawn Center (DAWN), which is funded by the Danish National Research Foundation under grant DNRF140.

We thank Eric F. Jim\'enez-Andrade for sharing their data products from the 10\,GHz \textit{Karl G. Jansky }Very Large Array survey of GOODS-N. The National Radio Astronomy Observatory is a facility of the National Science Foundation operated under cooperative agreement by Associated Universities, Inc. 

GL thanks Joris Witstok and Thomas Henning for enlightening discussions on dust production, and warmly thanks Fengwu Sun for sharing his Python code used to generate Figure 4.
\end{acknowledgements}

\bibliographystyle{aa} % style aa.bst
\bibliography{bibtex_goodsn} % your references Yourfile.bib

\begin{appendix}

\section{Dust production \label{App:dust_prod}}
\begin{figure}
\centering
\includegraphics[width=0.9\linewidth]{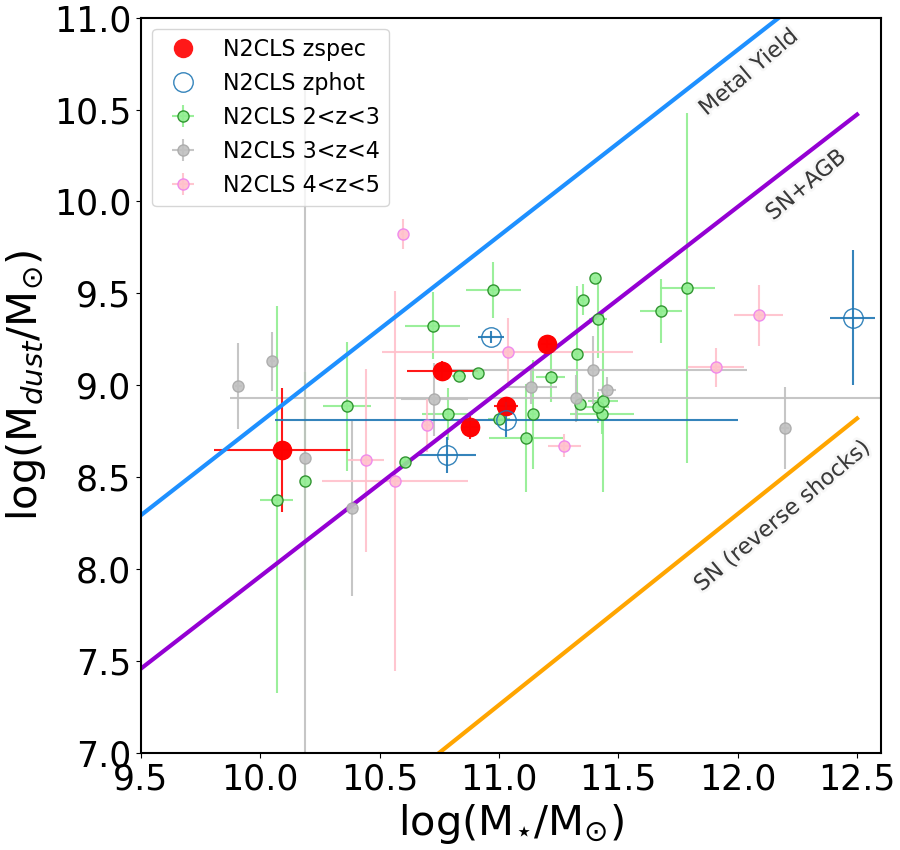}
\caption{Dust mass versus stellar mass for N2CLS galaxies. Filled red circles represent the five DSFGs with spectroscopic redshifts at $z \sim 5.2$, while open blue circles indicate those with photometric redshifts (including the very massive N2GN\_1\_20 galaxy discussed in Appendix\,\ref{App:ID20}). For comparison, we also show all N2CLS galaxies at $2 < z < 5$ from \citet{berta2025}. The region between the orange and purple lines is the expected dust mass range assuming a maximal dust production from both AGB stars and supernovae (SN), without significant reverse-shock destruction. The area below the orange line represents the same scenario, but accounting for $\sim$95\% reverse-shock destruction. The blue line shows the upper limit on dust mass based on the highest metal yields from \citet{witstok2023}, i.e. the extreme case of metal accretion onto dust grains resulting in maximal grain growth. \label{Fig:Dust_production}}
\end{figure}

Observations have shown a relation between dust mass and stellar mass across an extensive redshift range.  At all redshifts, the dust masses increase linearly with stellar masses. However, at fixed stellar mass, there is a large scatter of dust masses around the median, revealing the existence of a dust-poor galaxy population across all stellar masses \citep{Triani2020}. 
We compare in Fig.\,\ref{Fig:Dust_production} our stellar-mass estimates with our inferred dust masses for the eight $z\sim5.2$ DSFG. We also show the stellar mass–dust mass relations derived by \citet{witstok2023} based on dust production from stellar sources (including supernovae and AGB stars; with and without reverse shock destruction, which can lead to $\sim$95\% dust destruction). Additionally, we include the relation corresponding to the highest possible metal yield (i.e. assuming all metals are locked up in dust grains), representing the theoretical maximum dust mass.
Note that the exact normalisation of these relations remains uncertain due to systematic effects, including the choice of IMF used to compute dust yields for a given stellar mass, and the range of plausible dust absorption cross-sections.

The dust masses observed in our galaxies point to efficient dust formation. Indeed our galaxies are consistent with the dust masses expected from a simple, IMF-averaged yield of stellar sources alone (i.e. supernovae and AGB stars), assuming no dust destruction in the ISM or ejection from the galaxy. 
In general, it is challenging to populate our observed M$_\star$--M$_\mathrm{dust}$ relation using cosmological simulations, as our dusty galaxies occupy the upper envelope in both stellar and dust mass. They are, for instance, only sparsely represented in \cite{2023MNRAS.519.4632D} and entirely absent in the sample of \cite{Triani2020}.

We also show in Fig.\,\ref{Fig:Dust_production} the location of N2GN\_1\_20, the most massive galaxy in our z$\sim$5.2 sample, which is discussed in Appendix\,\ref{App:ID20}. It falls into the region where reverse-shock destruction of dust becomes significant, implying that additional dust growth mechanisms, such as grain growth in the ISM, may be necessary to account for their observed dust masses. However, we caution that this source currently lack spectroscopic redshift, and robust spectroscopic confirmation is essential before drawing firm conclusions about its dust production.

For comparison, we also present the complete sample of N2CLS galaxies at $2<z<5$ from \cite{berta2025}, which occupy the same region i.e. near the supernova+AGB dust production line. Notably, three galaxies lie above the maximum dust mass allowed: N2GN\_1\_17a ($z_{\text{spec}}$=3.19) and N2GN\_1\_32 ($z_{\text{spec}}$=3.65; detected in the Chandra 2 Ms map and requiring an AGN-torus component in the IRAC and MIPS bands) remain consistent within 1.3$\sigma$ (in M$_\mathrm{dust}$), while N2GN\_1\_04 ($z_{\text{spec}}$=4.056, also known as GN20) is significantly offset. 

The distribution of our measurements in the M$_\mathrm{dust}$--M$_\star$ plane appears relatively flat. \citet{Sommovigo2022} show that the M$_\mathrm{dust}$--M$_\star$ relation differs between $z\sim7$ REBELS galaxies and $z\sim5$ ALPINE galaxies (even when using consistent [CII]-based methods to estimate M$_\mathrm{dust}$), with REBELS galaxies exhibiting a flat trend, while ALPINE galaxies, as well as theoretical models (see Fig.\,\ref{Fig:Dust_production} and their Fig. 3), suggest that more massive galaxies are dustier. Both samples probe lower dust and stellar masses than our NIKA2 sources. The observed flatness of the M$_\mathrm{dust}$--M$_\star$ relation could partly arise from observational biases: the REBELS survey covers a limited stellar mass range, while for NIKA2, the 1.2\,mm flux selection imposes a dust mass detection threshold.

\section{Photometric redshifts with New-Hyperz \label{App_hyperz}}
For the four sources with literature-based photometric redshifts at $z\sim5.2$ (N2GN\_1\_20, N2GN\_1\_26, N2GN\_1\_43, N2GN\_1\_57), we re-derived their redshifts independently using the updated photometry and the New-Hyperz SED-fitting code. New-Hyperz is a modified  version  of  the  public code Hyperz originally developed by \citet{Bolzonella_2000}.
 The library of templates used for this need includes: (i) Eight evolutionary synthetic SEDs computed with the Bruzual \& Charlot code \citep{bc03}, with Chabrier IMF \citep{chabrier2003} and solar metallicity, matching the observed colours of local galaxies from E to Im types (namely a delta burst single stellar population, a constant star-forming system, and six $\tau$-models with exponentially decaying SFR); (ii) Six empirical templates for galaxies in the local Universe: four SEDs compiled by \cite{coleman1980}, and two starburst galaxies from the \cite{Kinney1996} library; (iii) Ten Starbursts99 evolutionary models \citep{STARBURST99}, including e-lines, for single bursts and constant SFR models, each spanning five metallicities (Z$=$ 0.04, 0.02, 0.008, 0.004 and 0.001). The explored redshift domain goes from z$=$0 to 16, without any prior in luminosity. The only restriction is that the age of the stellar population should be smaller than the age of the Universe for a given redshift. The internal dust extinction is a free parameter that is allowed to take values between A$_V=$0 and 6 magnitudes following the \citet{Calzetti2000} extinction law. During the SED-fitting process, to account in a realistic way for calibration uncertainties related to the combination of data coming from different facilities, across a wide range in wavelengths, we imposed a minimum photometric error of 0.2 magnitudes, which is quadratically added to the original error bars. Indeed, in our experience this is needed to prevent artificially small error bars leading to low-quality fits, by correcting from effects such as biases between different instruments due to (absolute) zero-point uncertainties, photometry computed with an effectively different spatial sampling and spatial resolution, and contribution to the total flux coming from different spatial regions. 

The quality of the photometric redshifts with New-Hyperz and the settings described above for the SED-fitting has been checked using the 29 N2CLS galaxies with spectroscopic redshifts (Sect.\,\ref{N2CLS}), excluding N2GN\_1\_61 because it is only detected in one filter. Spectroscopic redshifts are in the range between z$=$0.6 and 7.2. The results are as follows, with $\Delta z = z_{\rm phot} - z_{\rm spec}$:
$\sigma(\Delta z/(1+z_{\rm spec}))$ = 0.058, the median of $(\Delta z/(1+z_{\rm spec}))$ = -0.016, and the normalised median absolute deviation, which is less sensitive to outliers:
$\sigma_{z,MAD}$ = 1.48 $\times$ median $(|\Delta z|/(1+z_{\rm spec}))$ = 0.018. Regarding the fraction of outliers, defined in a conservative way as sources with $\Delta z >0.15(1+z_{\rm spec})$, it is found to be 21\% (6 of 28  objects), and only two galaxies (7\%) exhibit $|\Delta z|>0.30(1+z_{\rm spec})$. 
Among these six outliers, four exhibit a degenerate solution with a secondary solution consistent with $z_{\rm spec}$. One of them is an X-ray AGN at z$_{\rm spec}$ = 3.65 with a near-IR to mid-IR SED dominated by a torus component, exhibiting a degenerate solution between z$_{\rm phot}=0.085_{-0.085}^{+0.038}$ and z$_{\rm phot}=2.960_{-0.04}^{+0.11}$. %N2GN_1_32
Also, a poor z$_{\rm phot}$ estimate is obtained for an obscured quasar at z$_{\rm spec}$=7.19 \citep{Fujimoto2022}. 

\section{Notes about individual sources and JWST cutouts \label{App:notes}}

\begin{figure*}
\centering
\caption{JWST/NIRCam 10"$\times$10" cutout images centred on N2GN\_1\_01, 06, 13, 23, 43, and 61 (from top to bottom). The filters used are indicated at the top of each corresponding multi-colour image (on the right). White contours represent the NOEMA signal-to-noise ratio, ranging from 3$\sigma$ to 19$\sigma$ in steps of 2. N2GN\_1\_20, 26, and 57 are not covered by current JWST data.}
\label{Fig:JWST_stamps}
\includegraphics[width=0.67\linewidth]{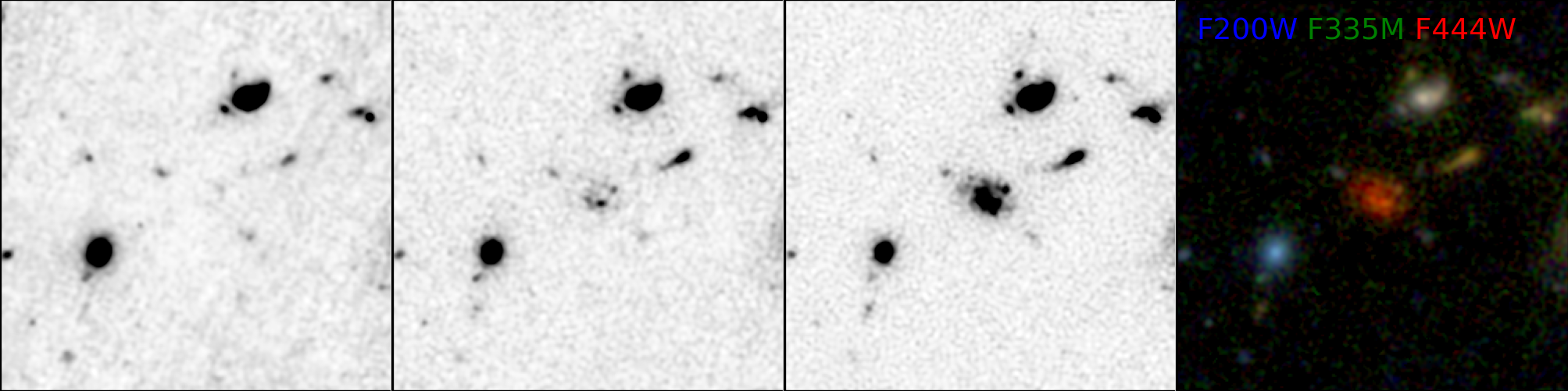}
\includegraphics[width=0.67\linewidth]{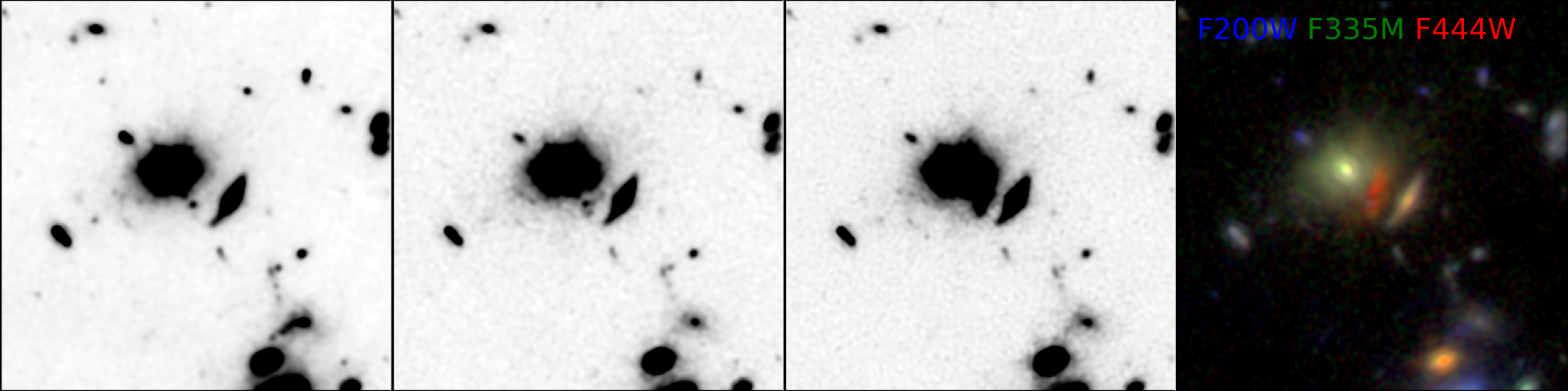}
\includegraphics[width=0.67\linewidth]{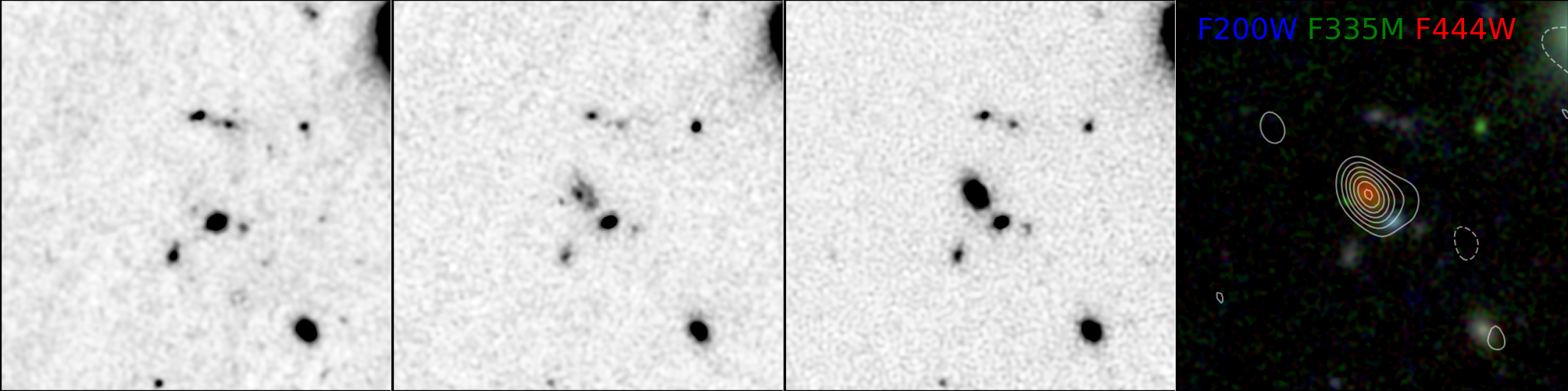}
\includegraphics[width=0.67\linewidth]{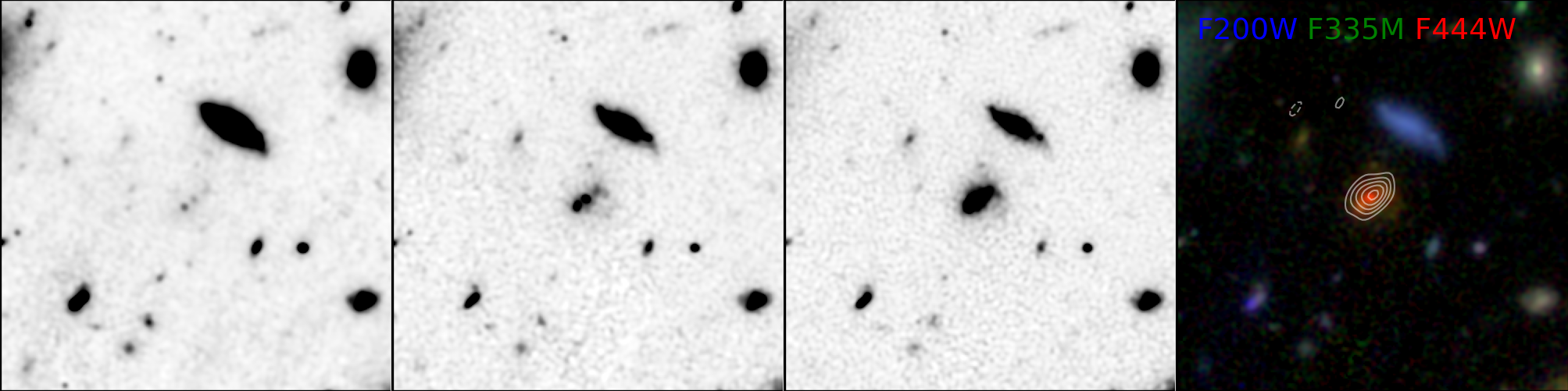}
\includegraphics[width=0.67\linewidth]{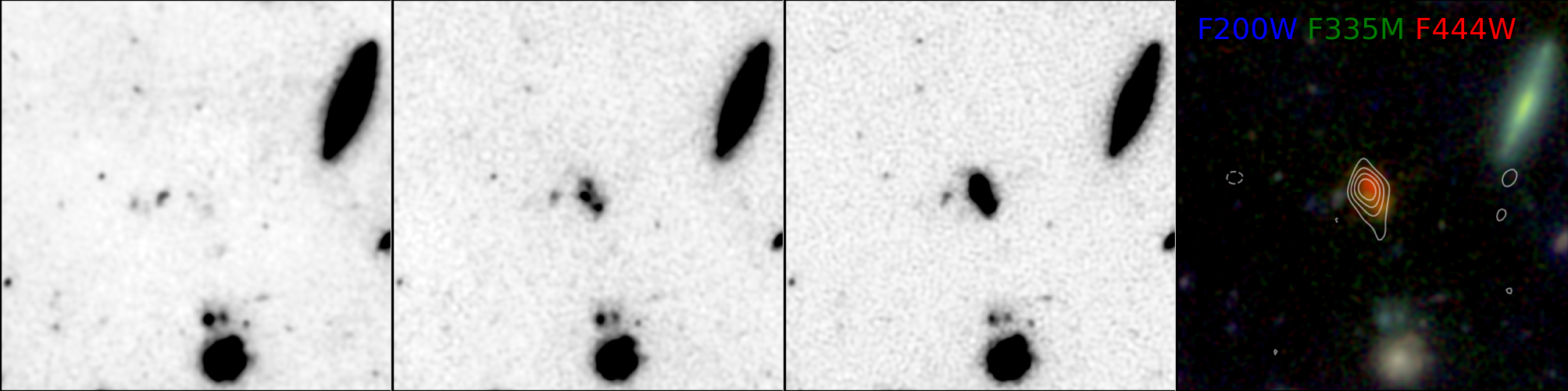}
\includegraphics[width=0.67\linewidth]{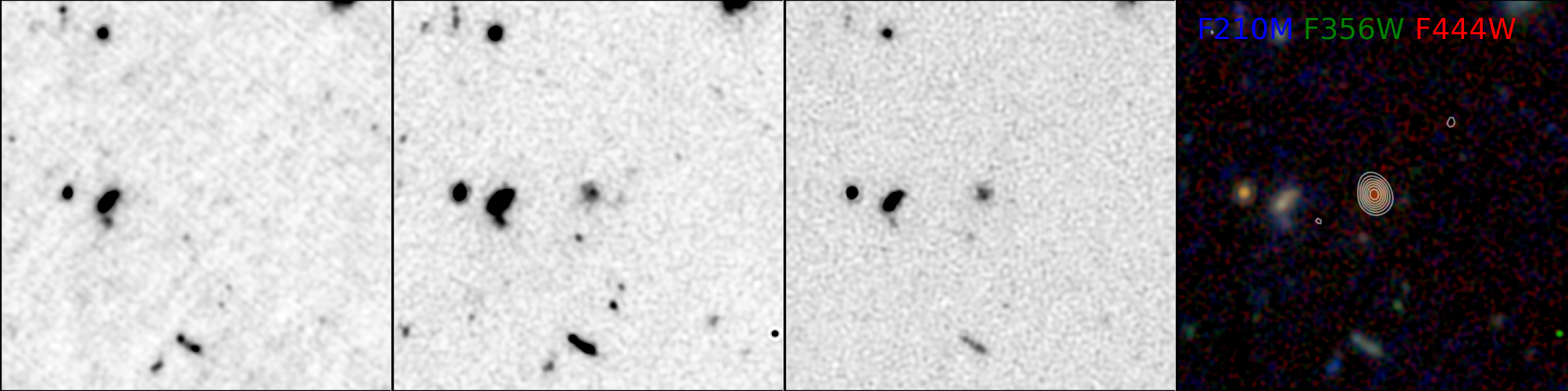}
\end{figure*}

Below, we provide a detailed description of each individual source identified at z$\sim$5.2 in the overdensity (but HDF850.1, GN10, and N2GN\_1\_13).
The JWST photometry measurements follow a similar approach to that described by \citet{weibel2024}, with some modifications as outlined in Sect.\,2.8 of B25.
The multi-wavelength cutouts of all sources are shown in B25. We complement those cutouts with the JWST postage stamps, where available, in Fig.\,\ref{Fig:JWST_stamps}.

\begin{itemize}
\item N2GN\_1\_23: is detected by JWST FRESCO at $z_\textrm{spec}$=5.179 and is the so-called S3 source extensively discussed in \cite{Xiao2024}. We have a firm identification of our N2CLS source through our NOEMA follow-up programme (see B25) with the NOEMA counterpart lying at a distance of 0.167\,arcsec. The NIKA2 source has both a SCUBA2 450 and a 850\,$\rm \mu$m counterpart but the NOEMA position lies at 3.6\,arcsec from the 450\,$\rm \mu$m source Number 78 of \cite{Barger2022}, identified with a galaxy at z=0.29 (see their Table\,2). The 850\,$\rm \mu$m source is identified with SMM123656621205 (Number 80), which is at 2.8\,arcsec from the NOEMA position, and with a redshift estimate of 3.75 \citep{Cowie2017}. The high-z source is clearly blended with a nearby object and flux measurements from Herschel and SCUBA2 450\,$\mu$m cannot be used to construct the multi-wavelength SED. The NOEMA source has both a 1.4\,GHz \citep{owen2018} and 10\,GHz \citep{jimenez2024} radio counterpart. 
\\
\item N2GN\_1\_26 is identified with a high-$z$ HST galaxy thanks to SMA data. \citet{kodra2023} give a photometric redshift $z_\textrm{phot}=5.31\pm0.08$. The far-IR SED is well defined by the PACS, SCUBA2, SMA, and NIKA2 data. The 24\,$\mu$m flux is well matched by an intense 3.3 $\mu$m polycyclic aromatic hydrocarbon feature. There is a 10\,GHz source at the HST position. 
This source is newly observed and detected by the PANORAMIC survey with F277W, F356W. Using New-Hyperz, it has a photometric redshift of 4.87$_{-0.19}^{+0.21}$, consistent with previous findings at 1$\sigma$ level. 
\\
\item  N2GN\_1\_43 is identified with NOEMA. It is a K-band dark galaxy. It is detected by JWST at $\sim 10\sigma$ in the F277W band, $\sim3\sigma$ in F200W, and tentatively by HST at shorter wavelengths. Although not in the IRAC catalogue, it is clearly seen in the maps at 4.5, 5.8 and 8\,$\mu$m. It has also a clear 10\,GHz signal \citep{jimenez2024} at the JWST source position.
A fit with CIGALE in B25 places it at $z_\textrm{phot}=5.87\pm0.38$ with a $\sim$4000\,\AA\, break between the two JWST measurements mentioned here. The photometric redshift from New-Hyperz is 5.31$_{-0.86}^{+1.33}$, consistent with previous estimates, with a degenerated solution in the parameter space between redshift and extinction leading to broad error bars.
\\
\item N2GN\_1\_57: The VLA source, coinciding with GN4\,26261 (MIOP programme) at z=0.9, is likely not the right counterpart as it lies at 4.9\,arcsec from the N2CLS position. Closer to the N2CLS position (2.2\,arcsec), the source HRG14 J123731.66+621616.7 \citep[also detected by the SHARDS middle-band survey;][]{finkelstein2015} is at a photometric redshift of z=5.2$\pm$0.06 in \cite{arrabal_haro2018} and z=5.30$_{-0.19}^{+017}$ in \cite{kodra2023}.
It is detected in the four IRAC channels and has a pronounced $\sim$4000\,\AA\ break at $\lambda\sim 2$\,$\mu$m. The photometric redshift obtained with New-Hyperz is z=5.18$_{-0.20}^{+ 0.19}$, consistent with previous findings. 
\\
\item  N2GN\_1\_61 is not only optically dark but also IRAC-dark ([4.5]>24.6\,mag at 5$\sigma$). The accurate position is given by NOEMA (see B25) revealing a single source with a flux of 698$\pm$84$\,\mu$Jy at 1.18\,mm, in perfect agreement with the NIKA2 flux. JWST grism observations from FRESCO detected one $\sim7\,\sigma$ emission line, consistent with H$\alpha$ at $z_\textrm{spec}=5.201$ (M.~Xiao, priv. comm.). It is observed by JWST with F090W, F115W, F182M, F210M, F356W, F444W filters, but only detected by F356W and F444W (see Fig.\,\ref{Fig:JWST_stamps}). 
The observed SED consists of only three data points (NIKA2 1.2\,mm, SCUBA2 850\,$\mu$m and JWST F444W), 
but the availability of the redshift allows us to perform a fit. Derived parameters have large uncertainties (but are consistent between MAGPHYS-highz and CIGALE). 
\end{itemize} 

\newpage
\section{Another N2CLS candidate at z$\sim$5.2 \label{App:ID20}}

We identified an additional N2CLS galaxy at z$\sim$5.2 that lies close to the overdense structure. N2GN\_1\_20 is detected by SMA. SMA position (RA=12:35:50.22 ; Dec=62:10:42.4) lies at 1.2\,arcsec from S-CANDELS J123550.35+621041.8, which is an optically dark galaxy. It has a Ks band flux of 0.75\,$\mu$Jy, resulting in a very red 3.6\,$\mu$m  to Ks colour ($S_{3.6}/S_\textrm{Ks}\sim13$), interpreted as a deep break at $\sim$4000\,\AA\, combined to a large extinction, that leads to a $z_\textrm{phot}=5.333\pm0.317$ \citep{liu2018}. Using New-Hyperz, we obtained $z_\textrm{phot}$=5.16$_{-1.10}^{+2.8}$, the broad error bars being the result of a degenerated solution in the parameter space between redshift, extinction and age of the stellar population. This source is not covered by JWST. 

\begin{figure}
\centering
\includegraphics[width=0.9\linewidth]{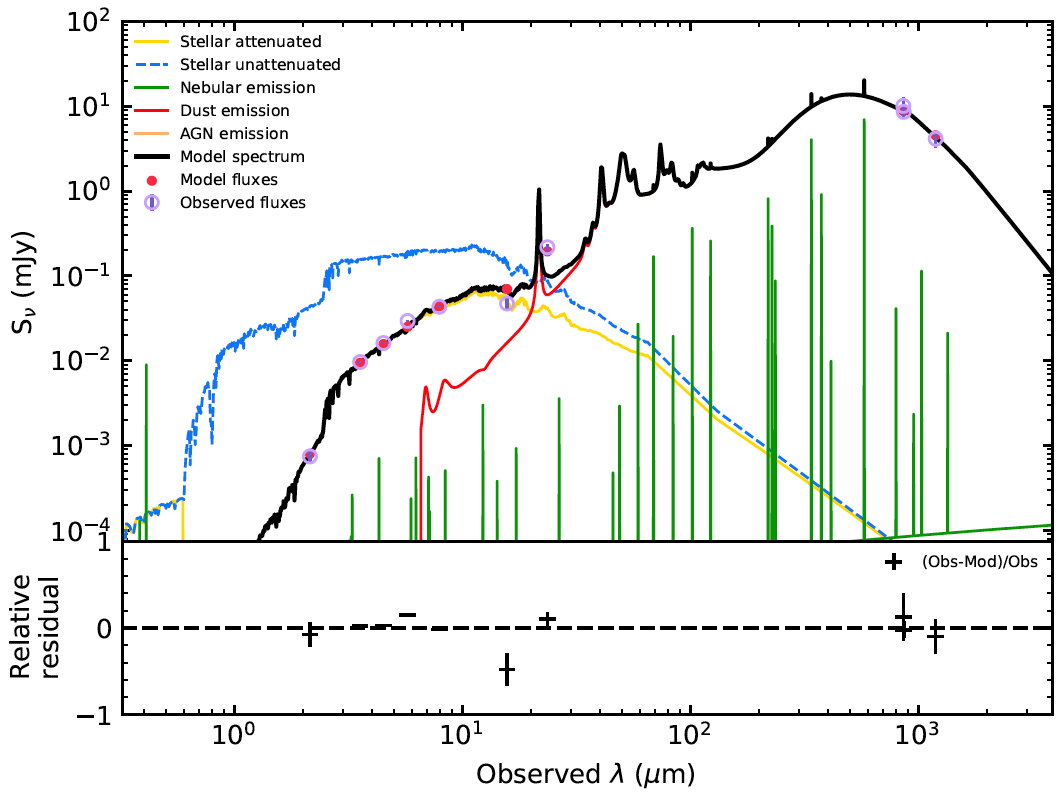}
\includegraphics[width=0.9\linewidth]{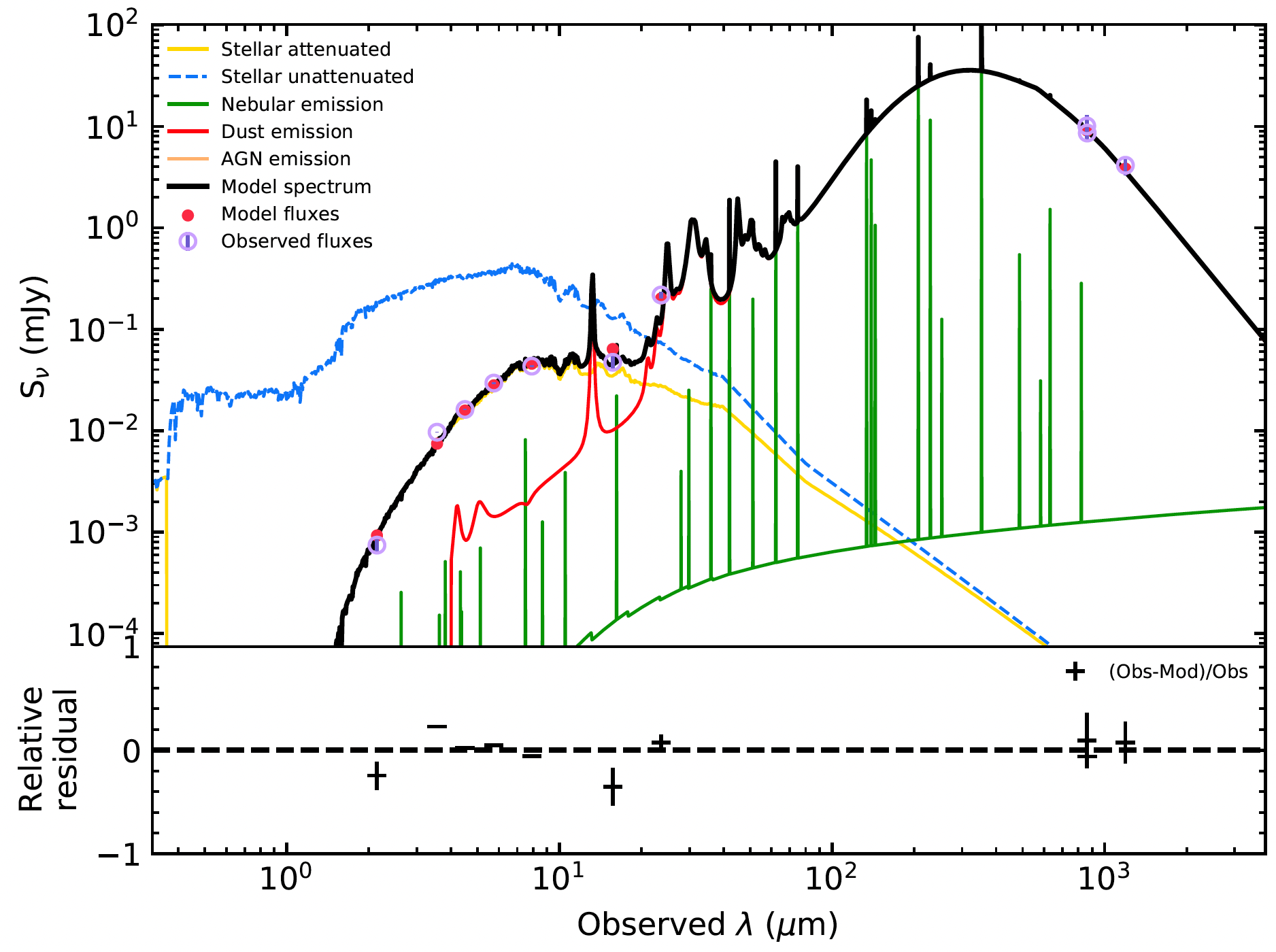}
\caption{N2GN\_1\_20: best fit from CIGALE assuming z=5.33 (top) and z=3 (bottom). Observed and modelled fluxes are shown with purple and red circles, respectively.}
\label{fig:cigale_ID20}
\end{figure}

We show in Fig.\,\ref{fig:cigale_ID20} the best fit from CIGALE, assuming z=5.33. The parameters derived from the fit are: SFR = 2160$\pm$611\,M$_{\odot}$\,yr$^{-1}$;
A$_{\rm V}$ = 2.7$\pm$0.3;
M$_{\star}$ = (303$\pm$66)$\times$10$^{10}$\,M$_{\odot}$; and M$_{\rm dust}$ = (23.3$\pm$19.70)$\times 10^{8}$\,M$_{\odot}$.
Combining  the  derived  SFR  and M$_{\star}$, and applying the \cite{tacconi2020} scaling relations, we obtain M$_{\rm gas}$= 103$\times$10$^{10}$\,M$_{\odot}$ and $\tau_{\rm dep}$ = 476\,Myr.
Assuming its photometric redshift, the galaxy is extremely massive. As such, it is lying between the \citet{popesso2023} and \citet{speagle2014} parametrisations of the MS. 

It is also lying above the model expectations for the GOODS-N survey volume (Fig.\,\ref{Fig:SFE}), requiring $\epsilon_{\star}$=0.7 for the full sky area. This discrepancy could arise from assumptions related to the IMF. For example, a top-heavy IMF could yield lower stellar masses by $\gtrsim 0.3$\,dex, resulting in slightly reduced values of $\epsilon_{\star}$.\ It could also arise from an overestimation of its photometric redshifts. A substantially lower true redshift compared to the photometric estimate may result in a lower best-fit stellar mass, thereby reducing the required value of $\epsilon_\star$. To test this hypothesis, we modelled the SED of N2GN\_1\_20 with CIGALE  assuming z=3 (see Fig.\,\ref{fig:cigale_ID20}). We obtain a stellar mass of (3.15$\pm$0.34)$\times10^{12}$\,M$_\odot$, essentially comparable to that inferred at z=5.33. It is important to note that the fit at z=3 is not merely a rescaling of the same template at a different redshift: CIGALE recomputes the fit and searches for an optimal solution at the imposed z. The reduced $\chi^2$ is 30\% higher at z=3 than at z=5.33. 
Such a high stellar mass at z=3 remains above model expectations for the GOODS-N field area, and would still require $\epsilon_{\star}$=0.12 for the full sky. 
If confirmed at z=5.33, the extremely high mass could reflect the presence of an unusually dense concentration of massive halos in this region.

\begin{figure}
\centering
\includegraphics[width=0.99\linewidth]{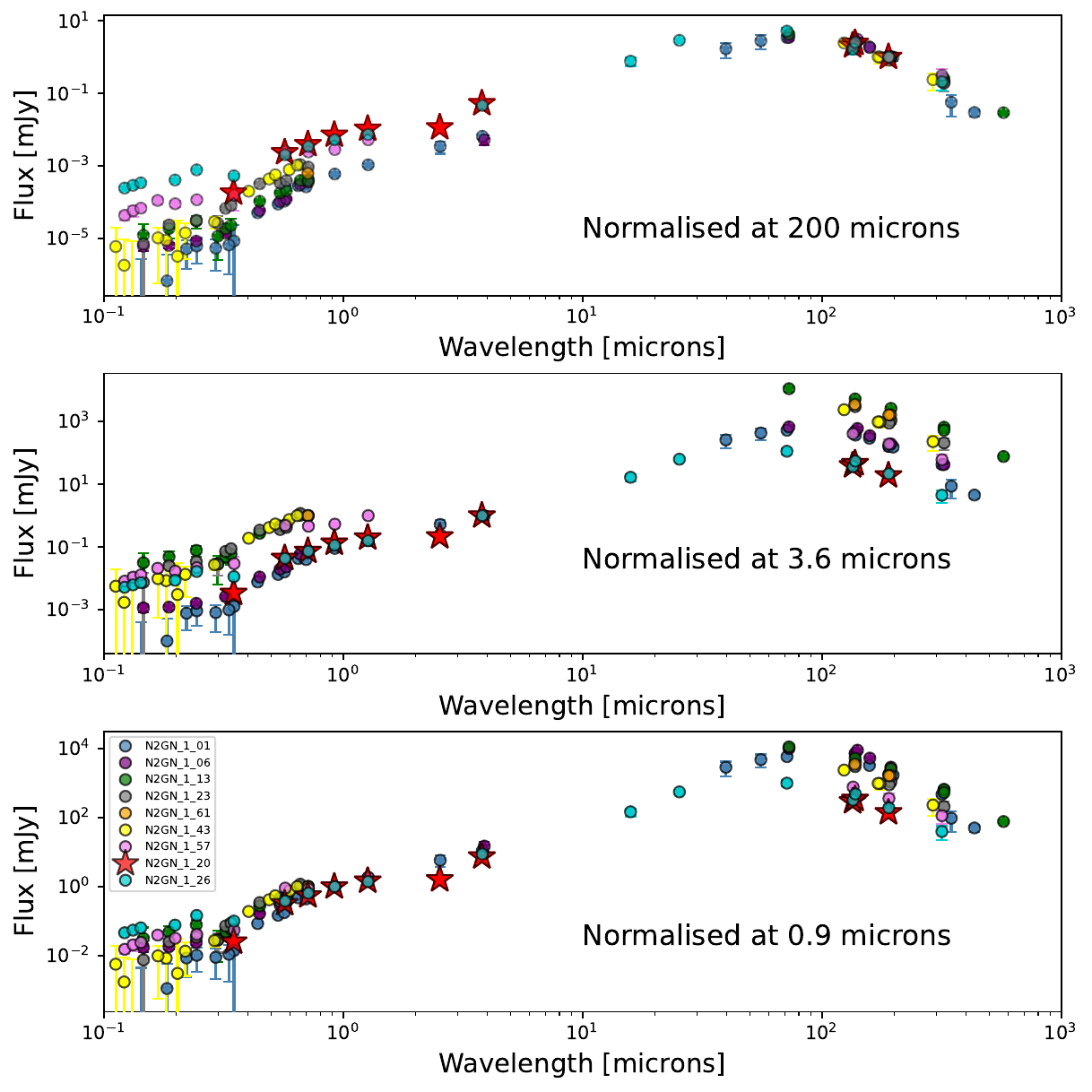}
\caption{SED of N2GN\_1\_20 at z=5.33 (red stars), compared to the eight N2CLS galaxies in the overdensity at z$\sim$5.2 (circles), shown as a function of rest-frame wavelength. From top to bottom, the SEDs are normalised at 200, 3.6, and 0.9\,$\mu$m (rest frame).}
\label{fig:SED_z5.pdf}
\end{figure}

We also compare the SED of N2GN\_1\_20 with those of the eight galaxies discussed in the main text, using different normalisations (Fig.\,\ref{fig:SED_z5.pdf}). We do not find any indication that N2GN\_1\_20 deviates significantly from the others.

As shown in Fig.\,\ref{Fig:3D}, N2GN\_1\_20 lies outside the core of the overdensity. Its distance to GN10 is $\Delta$RA=11.8\,cMpc, $\Delta$Dec=-8.1\,cMpc, and $\Delta$LOS$_\mathrm{distance}$= 14.9\,cMpc. Further investigation is required to determine whether this optically dark galaxy truly lies at high redshifts, to understand its unusually high stellar mass and efficiency, and to assess whether it may trace an extended region of the overdensity.

\end{appendix}

\end{document}